\documentclass[11pt,a4paper]{article}

\usepackage{amsmath,amssymb,tikz,hyperref,empheq}
\usepackage{graphicx}
 \usepackage{verbatim}
 \allowdisplaybreaks
\def\be{\begin{equation}}
\def\ee{\end{equation}}
\def\ba#1\ea{\begin{align}#1\end{align}}
\def\bg#1\eg{\begin{gather}#1\end{gather}}
\def\bm#1\em{\begin{multline}#1\end{multline}}
\def\bmd#1\emd{\begin{multlined}#1\end{multlined}}

\def\a{\alpha}
\def\b{\beta}

\def\d{\delta}

\def\e{\epsilon}

\def\({\left(}
\def\){\right)}
\def\[{\left[}
\def\]{\right]}

\def \be {\begin{equation}}
\def \ee {\end{equation}}
\def \ba {\begin{array}}
\def \ea {\end{array}}
\def \bea{\begin{eqnarray}}
\def \eea{\end{eqnarray}}

\def \a {\alpha}
\def \b {\beta}

\def \d {\delta}

\def \e {\epsilon}

\def\bea{\begin{eqnarray}}
\def\eea{\end{eqnarray}}
\newcommand{\eq}[1]{(\ref{#1})}
\newcommand{\bit}{\begin{itemize}}  \newcommand{\eit}{\end{itemize}}
\newcommand{\ben}{\begin{enumerate}}  \newcommand{\een}{\end{enumerate}}

\long\def\symbolfootnote[#1]#2{\begingroup%
\def\thefootnote{\fnsymbol{footnote}}\footnote[#1]{#2}\endgroup}

\newcommand{\sysu}{{\it School of Physics and Astronomy, Sun Yat-Sen University, 2 Daxue Road, Zhuhai 519082, China}}

\begin{document}
\thispagestyle{empty}
\begin{center}

~\vspace{20pt}

{\Large\bf Effective Action, Spectrum and First Law of Wedge Holography}

\vspace{25pt}

Peng-Ju Hu and Rong-Xin Miao ${}$\symbolfootnote[1]{Email:~\sf
  miaorx@mail.sysu.edu.cn}

\vspace{10pt}${}$\sysu

\vspace{2cm}

\begin{abstract}
In this paper, we study the effective action, the mass spectrum and the first law of entanglement entropy for a novel doubly holographic model called wedge holography. We work out the effective action of quantum gravity on the branes. In the perturbative formulation, it is given by an infinite sum of Pauli-Fierz actions. In the non-perturbative formulation, the effective action is composed of a higher derivative gravity and a matter action. Usually, a higher derivative gravity can be renormalizable but suffers the ghost problem. For our case, since the effective theory on the brane is equivalent to Einstein gravity in the bulk, it must be ghost-free. We notice that the matter action plays an important role in eliminating the ghost. We also provide evidences that the higher derivative gravity on the brane is equivalent to a ghost-free multi-gravity. Besides, we prove that the effective action yields the correct Weyl anomaly. Interestingly, although the effective action on the brane is an infinite tower of higher derivative gravity, the holographic Weyl anomaly is exactly the same as that of Einstein gravity. We also analyze the mass spectrum of wedge holography. Remarkably, there is always a massless mode of gravitons on the end-of-the-world branes in wedge holography. This happens because one imposes Neumann boundary condition on both branes. On the other hand, the massless mode disappears if one imposes Dirichlet boundary condition on one of the branes as in brane world theory and AdS/BCFT. Finally, we verify the first law of entanglement entropy for wedge holography. Interestingly, the massive fluctuations are irrelevant to  the first order perturbation of the holographic entanglement entropy.  Thus, in many aspects, the effective theory on the brane behaves like  massless Einstein gravity. 
\end{abstract}

\end{center}

\newpage
\setcounter{footnote}{0}
\setcounter{page}{1}

\tableofcontents

\section{Introduction}

Recently, a novel codimension two holography called wedge holography is proposed \cite{Akal:2020wfl}, which conjectures that the following dualities hold
\begin{eqnarray}\label{Wedgeholography}
\text{Classical gravity on wedge} \ W_{d+1}  &\simeq& \text{(Quantum) gravity on two branes}\  (Q_1\cup Q_2)  \nonumber\\
&\simeq &\text{CFT}_{d-1} \ \text{on corner of wedge}\  \Sigma , \nonumber
\end{eqnarray}
where $W_{d+1}$ is the $d+1$ dimensional wedge space, which is bounded by two end-of-the-world branes, i.e., $\partial W=Q_1\cup Q_2$, and $\Sigma$ is the corner of the wedge. See Fig.1 (left) for the  geometry. 
See also \cite{Bousso:2020kmy,Miao:2020oey,Miao:2021ual,Geng:2020fxl,Uhlemann:2021nhu,Uhlemann:2021itz} for some related works. Wedge holography is a generalization of the AdS/CFT correspondence \cite{Maldacena:1997re,Gubser:1998bc,Witten:1998qj}, and is closely related to brane world holography  \cite{Randall:1999ee,Randall:1999vf,Karch:2000ct}, AdS/BCFT \cite{Takayanagi:2011zk,Fujita:2011fp,Nozaki:2012qd,Miao:2018qkc,Miao:2017gyt,Chu:2017aab,Chu:2021mvq} and the doubly holographic model \cite{Penington:2019npb,Almheiri:2019psf,Almheiri:2019hni}.  
See
\cite{Rozali:2019day, Chen:2019uhq,Almheiri:2019psy,Kusuki:2019hcg,
   Balasubramanian:2020hfs,Geng:2020qvw,Chen:2020uac,Ling:2020laa,
   Kawabata:2021hac,Bhattacharya:2021jrn,Kawabata:2021vyo,
   Geng:2021hlu,Krishnan:2020fer,Neuenfeld:2021wbl, Neuenfeld:2021bsb,Chen:2020hmv,Ghosh:2021axl,Omiya:2021olc,Bhattacharya:2021nqj,Geng:2021mic,Sun:2021dfl,Chou:2021boq} for recent works on double holography and island. 
   As shown in Fig.1 (right),  wedge holography can be obtained from AdS/BCFT by taking the zero-volume limit $M\to 0$. In this limit, the bulk mode of CFTs on $M$ disappears and only the edge mode on the boundary $\Sigma$ survives. Thus wedge holography can be regarded as a holographic dual of the edge mode on the boundary (codim-1 defect) \cite{Miao:2021ual}.  Generalizing wedge holography to codim-n defects, \cite{Miao:2021ual} proposes the so-called cone holography, which can be derived from suitable limit of holographic defect CFT \cite{Jensen:2013lxa,DeWolfe:2001pq,Dong:2016fnf}.
 
The gravitational action of wedge holography is given by  \cite{Akal:2020wfl} 
\begin{eqnarray}\label{action}
  I_W=\frac{1}{16\pi G_N}\int_W \sqrt{|g|} (R-2\Lambda)
  +\frac{1}{8\pi G_N}\int_{Q_1 \cup Q_2} \sqrt{|h|} (K-T),
\end{eqnarray}
where $W$ denotes $d+1$ dimensional wedge space, $Q_1$ and $Q_2$ denote two end-of-the-world branes, $K$ is the trace of the extrinsic curvature, and $T$ is the tension of the brane. Following \cite{Takayanagi:2011zk}, \cite{Akal:2020wfl}  proposes to impose Neumann boundary condition (NBC) on the end-of-the-world branes
\begin{eqnarray}\label{NBC}
\text{NBC}:\ \left(K^{ij}-(K-T) h^{ij}\right)|_Q=0.
\end{eqnarray}
Actually, the Dirichlet boundary condition (DBC) \cite{Miao:2018qkc}
\begin{equation}\label{DBC}
\text{DBC}: \ \delta h_{ij}|_Q=0,
\end{equation}
and the conformal boundary condition (CBC) \cite{Chu:2021mvq}
\begin{subequations}\label{CBC}
  \begin{empheq}[left=\text{CBC}:\empheqlbrace]{align}
&
K=\frac{d}{d-1} T, \label{CBCa}\\
&  \delta h_{ij}|_Q= 2\sigma(y) h_{ij}|_Q, \label{CBCb}
  \end{empheq}
\end{subequations}
with $\sigma(y)$ a conformal factor also work well. See also \cite{Witten:2018lgb,Anderson:2006lqb,Anderson:2007jpe,Anderson:2010ph,York:1972sj,Papadimitriou:2005ii} for discussions on various boundary conditions of gravity. 
For simplicity, we mainly focus on NBC in this paper. 
It is found that wedge holography can yield the expected free energy, Weyl anomaly,  entanglement/R\'enyi entropy, two point functions and so on  \cite{Akal:2020wfl, Miao:2020oey}.  In particular, it obeys the holographic c-theorem  \cite{Miao:2020oey}. For one novel class of solutions, it is proved that wedge holography is equivalent to AdS/CFT with Einstein gravity \cite{Miao:2020oey}. These are all strong supports for wedge holography. 

The previous works \cite{Akal:2020wfl, Miao:2020oey} mainly focus on a special class of solutions. In this paper, we discuss the general solutions and gain more understanding of wedge holography. We work out the effective action of quantum gravity on the end-of-the-world branes with NBC. In the perturbative formulation, the effective action is given by an infinite sum of Pauli-Fierz actions of massive gravity. In the non-perturbative formulation, it is given by a higher derivative gravity plus a matter action. Usually, a higher derivative gravity suffers the problem of ghost. However, since the effective action on the brane is equivalent to Einstein gravity in the bulk, it must be ghost-free. We discuss the mechanism to eliminate the ghost and find that the matter action plays an important role. 
We argue that the higher derivative gravity on the brane is equivalent to a ghost-free multi-gravity. Besides, we prove that the effective action can produce the correct Weyl anomaly. 
Interestingly, although the effective action on the brane is an infinite tower of higher derivative gravity, the holographic Weyl anomaly is exactly the same as that of Einstein gravity.  We also study the mass spectrum of wedge holography. 
Interestingly, we find that there is a massless mode on the end-of-the-world branes if we impose NBC on both of the two branes. On the other hand, the massless mode disappears if one imposes DBC/CBC on one of the branes as in brane world holography and AdS/BCFT. Finally, we verify the first law of entanglement entropy and provide more supports for wedge holography.  

\begin{figure}[t]
\centering
\includegraphics[width=6.1cm]{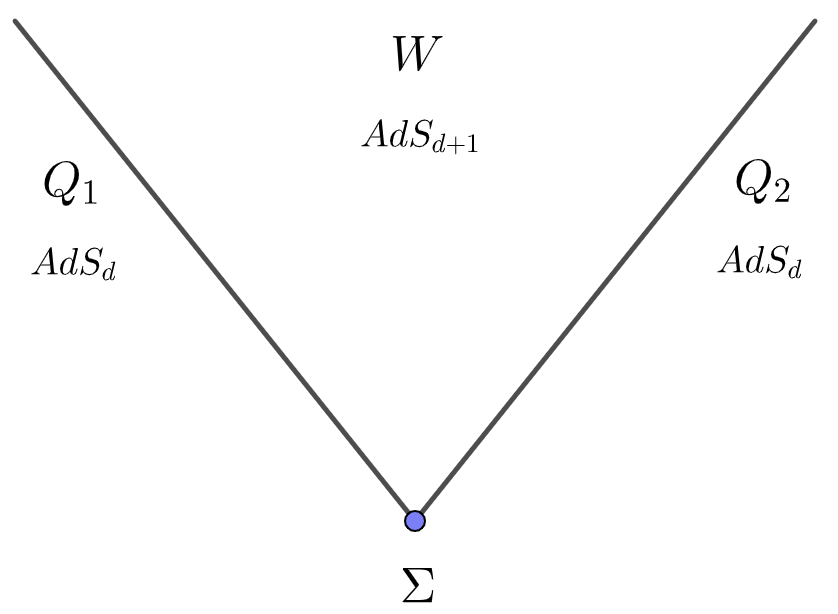}
\includegraphics[width=6.4cm]{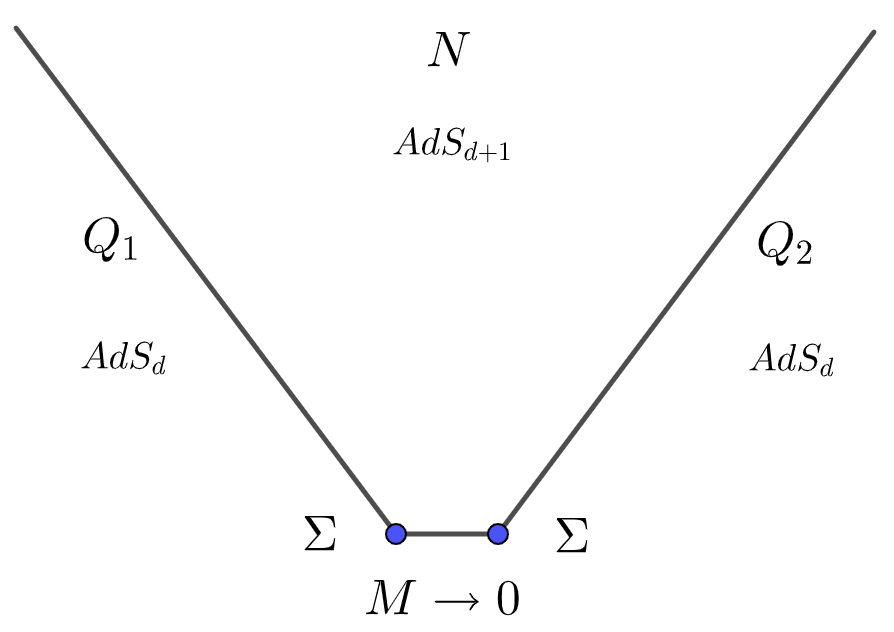}
\caption{(left) Geometry of wedge holography;\  \ (right) Wedge holography from AdS/BCFT.}
\label{wedge}
\end{figure}

The paper is organized as follows. 
In section 2, we investigate the effective action of gravity and discuss the mechanism to eliminate the ghost.  In section 3, we derive the holographic Weyl anomaly from the effective action of gravity on the brane. 
In section 4, we study the mass spectrum and perturbative effective action of vectors on the end-of-the-world branes.  In section 5, we generalize the discussions to the gravitons on the end-of-the-world branes. In section 6, we discuss the first law of entanglement entropy for wedge holography. Finally, we conclude with some open problems in section 6. 

Note added: It should be mentioned that, after this work has been finished, there appears an interesting paper \cite{Wang:2021xih} in arXiv, which also discusses the massless graviton on the brane.

\section{Effective action on the brane}

In this section, we investigate the effective action of wedge holography.  For simplicity, we focus on NBC. We leave the study of DBC/CBC to future works. We consider only gravity in this section. The generalization to the vector and the scalar is straightforward, but a little complicated.  

\subsection{Higher derivative gravity on the brane}
 
We apply the method of \cite{Chen:2020uac}, which is initially developed for the case of large brane tension $T=(d-1)\tanh(\rho)\to (d-1)$, where the end-of-the-world brane approaches to the AdS boundary. In fact, the method of \cite{Chen:2020uac} can be generalized to arbitrary brane tension. See \cite{Emparan:2020znc} for discussions of the case $d=3$. 

To warm up, let us first consider the case of large brane tension, i.e.,  $\rho\to \infty$ ($T\to d-1$), which means that  the brane is set at infinity and the geometry of wedge holography becomes that of AdS/CFT.  In this limit, the bulk gravitational  action $I_W$ (\ref{action}) is divergent. To get a finite action, one can perform the holographic renormalization by adding the following counterterms on the brane 
\cite{Balasubramanian:1999re,deHaro:2000vlm}
 \begin{eqnarray}\label{gravityIbdy}
&&I_{c}=\frac{-1}{16 \pi G_N}\int_Q \sqrt{|h|} \Big[2(d-1)-2T+\frac{1}{d-2} \mathcal{R}+\frac{1}{(d-4)(d-2)^2}\left(\mathcal{R}^{ij}\mathcal{R}_{ij}-\frac{d}{4(d-1)}\mathcal{R}^2\right)\nonumber\\
&&\ \ \ \ \ \ \ \ \ \ \ \ \ \ \ \ \ \ \ \ \ \ \ \ \ \ \ \ +\mathcal{L}_{\mathcal{R}3}+...\Big],
 \end{eqnarray}
 where $Q$ denotes $Q_1\cup Q_2$, $T=(d-1) \tanh(\rho)$ \footnote{We add T in the counterterm eq.(\ref{gravityIbdy}) in order to compensate the brane action 
 $-\frac{1}{8\pi G_N}\int_Q \sqrt{|h|}T$ in the bulk action eq.(\ref{action}). In the standard holographic renormalization, we have $T=0$ due to the absence of brane action.}, $\mathcal{R}_{ij}$ denotes the intrinsic curvature on the brane and $\mathcal{L}_{\mathcal{R}3}$ is given by \cite{Kraus:1999di}
 \begin{eqnarray}\label{Ic3}
&&\mathcal{L}_{\mathcal{R}3}=\frac{-2}{(d-2)^3(d-4)(d-6)}\Big( \frac{3d+2}{4(d-1)} \mathcal{R} \mathcal{R}_{ij} \mathcal{R}^{ij}-\frac{d(d+2)}{16(d-1)^2}  \mathcal{R}^3-2 \mathcal{R}^{ij}\mathcal{R}_{ikjl}  \mathcal{R}^{kl}\nonumber\\
&&\ \ \ \ \ \ \ \ \ \ \ \ \ \  \ \ \ \ \ \ \ \ \ \ \ \ \ \ \ \ \ \ \ \ \ \ \ \ \ \   +\frac{d}{4(d-1)} \mathcal{R}  \Box  \mathcal{R} - \mathcal{R}^{ij}\Box \mathcal{R}_{ij}\Big).
\end{eqnarray}
Now the renormalized action (the effective action of CFTs)
  \begin{eqnarray}\label{ICFT}
I_{\text{CFT}}=I_W+ I_c,
 \end{eqnarray}
becomes finite.  From eq.(\ref{ICFT}), we read off the action of wedge holography
  \begin{eqnarray}\label{IW}
I_W=I_{\text{CFT}}-I_c,
 \end{eqnarray}
which is composed of a matter part $I_{\text{CFT}}$ and a gravitational part $(-I_{c})$.  

Now set the brane at a finite place $r=\pm \rho$ instead of at infinity $r=\pm \infty$. Then $I_W$ becomes finite.  One can always separate $I_W$ into a matter part and a gravitational part.  The natural conjecture is that the gravitational part is still given by $(-I_{c})$ eq.(\ref{gravityIbdy}) even if the brane is located at a finite position.  One support for this proposal is that $(-I_{c})$ eq.(\ref{gravityIbdy}) yields the correct leading terms of entanglement entropy \cite{Chen:2020uac}.  Let us explain more on this point.  In double holography, one has two methods to calculate the entanglement entropy. One method is by using holographic entanglement entropy of higher derivative gravity $(-I_{c})$ eq.(\ref{gravityIbdy}) on the brane \cite{Dong:2013qoa,Camps,Miao:2014nxa}, the other way is by applying TR formula of Einstein gravity in the bulk \cite{Ryu:2006bv}. It turns out that these two methods yield the same entanglement entropy for large but finite brane tension \cite{Chen:2020uac}.  Another support for this proposal is that, as we will show in sect.3,  $(-I_{c})$ eq.(\ref{gravityIbdy}) yields the correct Weyl anomaly for wedge holography. 

Note that $I_c$  eq.(\ref{gravityIbdy}) includes infinite higher derivative terms for the finite tension $T< (d-1)$.  Note also that eq.(\ref{gravityIbdy}) works well only for odd $d$. For even $d$, one needs to correct  $I_{c}$ eq.(\ref{gravityIbdy}). See \cite{Chen:2020uac} for some examples.  For simplicity, we focus on the case of odd $d$ in this section. 

To end this subsection, let us give a further evidence for the proposal eq.(\ref{IW}). We verify that eq.(\ref{IW}) yields the correct effective action for the following type of solutions
  \begin{eqnarray}\label{bulkmetric123}
ds^2=dr^2+\cosh^2(r) \bar{h}_{ij} dy^i dy^i, 
 \end{eqnarray}
where $-\rho\le r\le \rho$, and $\bar{h}_{ij} $ obeys the vacuum Einstein equations on the brane
  \begin{eqnarray}\label{EinsteinEQonbrane}
 R_{\bar{h}\ ij}-\frac{R_{\bar{h}}+(d-1)(d-2)}{2} \bar{h}_{ij}=0.
 \end{eqnarray}
Here $R_{\bar{h}\ ij}$ is the curvature with respect to $\bar{h}_{ij}$. The effective action of wedge holography for the solution eq.(\ref{bulkmetric123}) is given by \cite{Miao:2020oey}
  \begin{eqnarray}\label{IWEinstein}
I_{\text{W Ein}}=\frac{1}{16\pi G_N}\int_{0}^{\rho}\cosh^{d-2}(r) dr \int_Q dy^d
\sqrt{|\bar{h}|} \Big(R_{\bar{h}}+(d-1)(d-2) \Big),
 \end{eqnarray}
 where we focus on half of the wedge space, i.e, $0\le r\le \rho$, for simplicity.
Rewriting the above action in form of the induced metric $h_{ij}=\cosh^2(\rho)\bar{h}_{ij}$ on the brane, we get
  \begin{eqnarray}\label{IWEinstein1}
I_{\text{W Ein}}=\frac{1}{16\pi G_N}\int_{0}^{\rho}\frac{\cosh^{d-2}(r)}{\cosh^{d-2}(\rho)} dr \int_Q dy^d 
\sqrt{|h|} \Big(\mathcal{R}+\frac{(d-1)(d-2)}{\cosh^2(\rho)} \Big).
 \end{eqnarray}
 For the class of solution eq.(\ref{bulkmetric123}), we have 
   \begin{eqnarray}\label{formualaction}
\mathcal{R}_{ij}=-(d-1)\text{sech}^2(\rho) h_{ij},\ \  \mathcal{R}=-d(d-1)\text{sech}^2(\rho).
 \end{eqnarray}
 By applying the above formula, eq.(\ref{IWEinstein1}) can be further simplified as
   \begin{eqnarray}\label{IWEinstein2}
I_{\text{W Ein}}&=&\frac{1}{16\pi G_N} \int_Q dy^d
 \sqrt{|h|} (d-1) \text{sech}^{d}(\rho ) B_{\text{sech}^2(\rho )}\left(1-\frac{d}{2},\frac{1}{2}\right)\nonumber\\
&=&\frac{1}{16\pi G_N} \int_Q dy^d 
\sqrt{|h|} \Big(-\frac{2 (d-1) \epsilon ^2}{d-2}+\frac{(1-d) \epsilon ^4}{d-4}-\frac{3 (d-1) \epsilon ^6}{4 (d-6)}+O\left(\epsilon ^8\right) \Big),
 \end{eqnarray}
 for odd $d$, where $B_a(b,c)$ denotes the beta function and we have reparameterized $\rho$ by
   \begin{eqnarray}\label{elable}
\e=\text{sech}(\rho).
 \end{eqnarray}
 
Note that the $\mathcal{R}^n$ terms in the counterterm eq.(\ref{gravityIbdy}) are of orders $O(\e^{2n})$. Substituting eqs.(\ref{formualaction},\ref{elable}) into the effective action eq.(\ref{IW}), we derive 
  \begin{eqnarray}\label{IW1}
I_W=I_{\text{CFT}}+\frac{1}{16\pi G_N} \int_Q dy^d
 \sqrt{|h|} \Big(-\frac{2 (d-1) \epsilon ^2}{d-2}+\frac{(1-d) \epsilon ^4}{d-4}-\frac{3 (d-1) \epsilon ^6}{4 (d-6)}+O\left(\epsilon ^8\right) \Big),
 \end{eqnarray}
which agrees with eq.(\ref{IWEinstein2}) provided that $I_{\text{CFT}}=0$.  The vanishing of the CFT action has a natural explanation in odd dimensions. In the limit $\rho\to \infty (\e\to 0)$, the geometry of wedge holography becomes that of AdS/CFT. We can check by direct calculations that  $I_{\text{CFT}}$ eq.(\ref{ICFT}) indeed vanishes for the class of solutions eq.(\ref{bulkmetric123}) when $d$ is odd. It is natural that $I_{\text{CFT}}$ remains zero as we adiabatically pull the brane from infinity to a finite place.  $I_{\text{CFT}}=0$ means that eq.(\ref{IWEinstein1}) is the vacuum solution to AdS/CFT, and 
the complete expressions of the counterterms $(-I_c)$ eq.(\ref{gravityIbdy}) is given by eq.(\ref{IWEinstein1}) for the solutions eq.(\ref{bulkmetric123}).  Let us go on to discuss the case of finite tension. For simplicity, we assume that the geometry on the brane is an AdS. Recall that the vacuum energy of CFTs vanishes in odd AdS space, since there is no Weyl anomaly in odd dimensions \footnote{One the other hand, the vacuum energy is non-zero for CFTs in even AdS space, due to the non-trivial Weyl anomaly. }. As a result, it is natural that the effective action of CFTs also vanishes for odd $d$.  

Now we finish the verification that $I_W$ eq.(\ref{IW}) gives the correct effective action of wedge holography for the class of solutions eq.(\ref{bulkmetric123}). As a by-product, we notice that, $I_{\text{CFT}}=0$ and the complete expression of  counterterms $(-I_{c})$ eq.(\ref{gravityIbdy}) is given by eq.(\ref{IWEinstein1}) for the solutions eq.(\ref{bulkmetric123}).

\subsection{Mechanism to eliminate ghosts}

As we have shown in the above subsection, the effective action on the end-of-the-world brane is an infinite tower of higher derivative gravity. Usually, a higher derivative gravity suffers the ghost problem. However, since the effective action is obtained from Einstein gravity in the bulk, it must be ghost-free.  In this subsection, we discuss the mechanism to eliminate the ghost and find that the matter action $I_{\text{CFT}}$ plays an important role. We also argue that higher derivative gravity on the brane is equivalent to a ghost-free multi-gravity. 

Let us first give a quick review of the ghost problem of higher derivative gravity. 
 Take  curvature squared gravity as an example. In general, 
 it includes a scalar mode, a massless graviton and a massive graviton.  
 Consider the linear perturbation equation around a flat-space background
\begin{eqnarray}\label{perEOMR2}
\Box \left(\Box-m^2\right)\delta h_{ij}=0,
\end{eqnarray}
where $\Box $ is the D'Alembert operator and the metric perturbation $\delta h_{ij}$ obeys $D_i  \delta h^{ij}=0$ and $h^{ij} \delta h^{ij}=0$. From eq.(\ref{perEOMR2}), we read off the propagator
\begin{eqnarray}\label{propagator}
D (p) \sim  \Big( \frac{1}{p^2}- \frac{1}{p^2+m^2}\Big),
\end{eqnarray}
where we have ignored the indexes and the tensor structures.  From eq.(\ref{propagator}), it is clear that the massless graviton and the massive graviton cannot both have the correct sign. In other words, one of them is a ghost. Our case is similar. The linear perturbation equation of the higher derivative gravity on the brane takes the form
\begin{eqnarray}\label{perEOMRinfnite}
\Pi_{n=0}^{\infty} (\Box+\frac{2}{L_{\text{eff}}^2}-m_n^2)\delta h_{ij}=0,
\end{eqnarray}
where $m_0=0$ and  $m_n$ denotes the mass of the nth mode. Without loss of generality, we set $m_0^2< m_1^2<m_2^2<...$. For the large AdS radius $L_{\text{eff}}=\cosh(\rho)\to \infty$, eq.(\ref{perEOMRinfnite}) yields the following propagator
\begin{eqnarray}\label{propagator2}
&&D(p) \sim \sum_{i=0}^{\infty} \Big(\Pi_{j\ne i} \frac{1}{m_j^2-m_i^2} \Big)\frac{1}{p^2+m_i^2} \nonumber\\
&&\ \ \ \ \ \ \  \sim \Big( \frac{a^2_0}{p^2}- \frac{a_1^2}{p^2+m_1^2}+ \frac{a_2^2}{p^2+m_2^2}- \frac{a_3^2}{p^2+m_3^2}+...\Big),
\end{eqnarray}
where $a_i^2$ are some positive constants related to the mass of gravitons. The above propagator
 implies that half of the massive modes are ghosts.  However, this contradicts with 
 the fact the effective action on the brane must be ghost-free, since it is obtained from Einstein gravity in the bulk. 

Let us discuss the resolution to the above puzzle. To start, we want to mention an enlightening example. It is found in \cite{Hassan:2013pca} that,  by eliminating one of the metric, the ghost-free bimetric theory \cite{Hassan:2011zd,Hassan:2011ea}
\begin{eqnarray}\label{bimetricgravity}
S[h, f]=m_{h}^{d-2}\int dy^d \left( \sqrt{|h|} \mathcal{R}(h)+\alpha^{d-2}\sqrt{|f|} \mathcal{R}(f) -2m^2\sqrt{|h|}\sum_{n=0}^d \beta_n e_n (S)\right),
\end{eqnarray}
is equivalent to the higher derivative gravity
\begin{eqnarray}\label{biHDgravity}
S^{HD}=m_{h}^{d-2}\int dy^d  \sqrt{|h|} \left( \Lambda+ c_R \mathcal{R}(h)-\frac{c_{RR}}{m^2}\left(\mathcal{R}^{ij}\mathcal{R}_{ij}-\frac{d}{4(d-1)}\mathcal{R}^2\right)+ O(\frac{\mathcal{R}^3}{m^4})\right),
\end{eqnarray}
where $\alpha=m_f/m_h$ is the ratio of the Planck masses, $\beta_n$ are dimensionless free parameters, $m^2$ sets the mass scale of the massive mode, $ \Lambda, c_R ,c_{RR} $ are some unimportant constants depending on the parameters of bimetric  gravity,  $S^i_j=(\sqrt{h^{-1} f})^i_j$ and $e_n(S)$ is given by
\begin{eqnarray}\label{bien}
e_n(S)=-\frac{1}{n} \sum_{k=1}^n(-1)^k \text{Tr}(S^k) e_{n-k}(S).
\end{eqnarray} 
Due to the equivalence to the ghost-free bimetric theory, the higher derivative gravity eq.(\ref{biHDgravity}) must be ghost-free too. However, as we have reviewed above, the curvature squared gravity includes a ghost. \cite{Hassan:2013pca} argues that the ghost problem is an artifact of the truncation at $O(\mathcal{R}^2)$. If one considers the complete action with infinite higher derivative terms, the ghost can be removed. This is indeed the case in a toy model of higher derivative scalar theory with suitable couplings to the source \cite{Hassan:2013pca}. It is found that the source structure plays an important role in eliminating the ghost in the higher derivative theory \cite{Hassan:2013pca}.

Now turn to our case. Remarkably, the curvature squared term of eq.(\ref{biHDgravity}) takes exactly the same form as that of the higher derivative gravity $I_W$ eqs.(\ref{gravityIbdy},\ref{IW}) on the brane.  Following the approach of \cite{Hassan:2013pca}, we find that, by choosing suitable parameters of the bimetric gravity, the $O(\mathcal{R}^3)$ term eq.(\ref{Ic3}) of $I_W$ can also be recovered \cite{HuandMiao}. This strongly implies that the higher  derivative gravity on the brane is actually ghost-free. It should be mentioned that, although the  $O(\mathcal{R}^2)$, $O(\mathcal{R}^3)$ terms are the same, the effective theory on the brane cannot be equivalent to the bimetric gravity eq.(\ref{bimetricgravity}). That is because they have different degrees of freedom. There are infinite massive modes on the brane, while there are only a massless mode and a massive mode in bimetric gravity.  The inconsistency of degrees of freedom can be naturally resolved by considering multi-gravity with infinite metrics \cite{Hinterbichler:2012cn,deRham:2014zqa}
\begin{eqnarray}\label{multi-HDgravity}
S_N=\lim_{N\to \infty}\frac{M^{d-2}}{2}  \sum_{n=1}^{N}\int dy^d  \sqrt{|h_n|} \left( \mathcal{R}_n +\frac{m_N^2}{2} \sum_{m=0}^{d}\alpha_m^{(n)} \mathcal{L}_m\left(\mathcal{K}(h_n, h_{n+1})\right)\right),
\end{eqnarray}
where $h_n$ denotes the nth metric, $\mathcal{R}_n$ is the Ricci scalar of $h_n$ and $ \mathcal{L}_m\left(\mathcal{K}(h_n, h_{n+1})\right)$ is the interaction between neighboring metrics.  Naturally, $h_1$ and $h_N$ are the metrics on the two branes. Following the approach of \cite{Hassan:2013pca}, we eliminate the other metrics and obtain two higher derivative gravity on the two branes
\begin{align}\label{multi-HDgravity}
S_N^{HD}&=\frac{M^{d-2}}{2}  \int dy^d \sqrt{|h_1|} \left( \Lambda_1+c_R   \mathcal{R}_1-\frac{c_{RR}}{m_N^2} \left(\mathcal{R}_1^{\;ij}\mathcal{R}_{1\ ij}-\frac{d}{4(d-1)}\mathcal{R}_1^2\right)+ O(\frac{\mathcal{R}_1^3}{m_N^4})\right)\nonumber\\
&+\frac{M^{d-2}}{2}  \int dy^d \sqrt{|h_N|} \left( \Lambda_N +c_R   \mathcal{R}_N-\frac{c_{RR}}{m_N^2} \left(\mathcal{R}_N^{\;ij}\mathcal{R}_{N\ ij}-\frac{d}{4(d-1)}\mathcal{R}_N^2\right)+ O(\frac{\mathcal{R}_N^3}{m_N^4})\right)\nonumber\\
&+ \frac{M^{d-2}}{2}  \int dy^d   \mathcal{L}_{int}(h_1, h_N),
\end{align}
where $\mathcal{L}_{int}(h_1, h_N)$ denotes the interaction of $(h_1, h_N)$. See the appendix for the derivations. Remarkably, the $O(\mathcal{R}^2)$ term takes exactly the same form as that of the effective action $I_W$ eqs.(\ref{gravityIbdy},\ref{IW}).  This is a strong evidence that the effective theory on the brane is equivalent to a ghost-free multi-gravity.  We leave a careful study of this equivalence to future work \cite{HuandMiao}. 

To end this section, we provide an alternative mechanism to remove the ghost in higher derivative gravity. 
To have a ghost-free unitary theory, we require that the Euclidean effective action satisfies the following condition
\begin{eqnarray}\label{unitary}
e^{-I_W}=e^{-I_{\text{CFT}}+I_c} \le 1. 
\end{eqnarray}  
Due to the ghost, the exponent of higher derivative action $e^{I_c}$ could be larger than one.  Thus, the matter action $I_{\text{CFT}}$ plays an important role in defining a unitary theory on the brane.  Indeed, $I_{\text{CFT}}$ is large enough to make $e^{-I_W}$ less than one. That is because the effective action $I_W$ on the brane is obtained from the Einstein gravity in the bulk. As a result, it must satisfy the above condition
\begin{eqnarray}\label{unitary}
e^{-I_W}=e^{-I_{\text{Ein bulk}}} \le 1. 
\end{eqnarray} 
This teaches us an interesting mechanism to eliminate the ghost.  By adding suitable matter fields to the higher derivative gravity, it is possible to construct a ghost-free total theory.  As we have discussed above, wedge holography provides such an example. However, the effective action of wedge holography includes infinite higher derivative terms, thus is too complicated. Is it possible to construct a ghost-free total theory with finite higher derivative terms?  We hope  this interesting problem could be addressed in futures.

\section{Holographic Weyl anomaly}

In this section, we derive the holographic Weyl anomaly from the effective action eq.(\ref{IW}) of wedge holography and show that it agrees with the result of \cite{Miao:2020oey}. This can be regarded as a test of the effective action eq.(\ref{IW}).

According to \cite{Henningson:1998gx}, the Weyl anomaly \cite{Weylanomaly} can be obtained from the UV logarithmic divergent term of the gravitational action.  Since $I_{\text{CFT}}$ is finite, it is expected that it does not contribute to the UV logarithmic divergent term. Thus, we focus on the gravitational part $(-I_c)$ of the effective action eq.(\ref{IW}). We assume that the spacetime on the brane is asymptotically AdS
\begin{eqnarray}\label{metricanomaly}
ds_Q^2=h_{ij}dy^i dy^j=\cosh^2(\rho) \frac{dz^2+\sigma_{ab} dy^a dy^b}{z^2},
\end{eqnarray}
where $y^i=(z, y^a)$, $\sigma_{ab} =\sigma^{(0)}_{ab} + z^2 \sigma^{(1)}_{ab} +...+z^{d-1} ( \sigma^{(\frac{d-1}{2})}_{ab}+\lambda^{(\frac{d-1}{2})}_{ab}\ln z)$.  By analyzing the asymptotical symmetry of AdS,  \cite{Imbimbo:1999bj} obtains a universal relation for general higher derivative gravity
\begin{eqnarray}\label{g12d}
&& \sigma^{(0)ab}\sigma^{(1)}_{ab}=-\frac{R_{\Sigma}}{2},  \ \ \ \ \ \ \ \ \ \ \ \ \ \ \ \ \ \ \ \ \ \ \ \ \text{for} \ d=3 \\
&& \sigma^{(1)}_{ab}=\frac{-1}{d-3}(R_{\Sigma\ ab}-\frac{R_{\Sigma}}{2(d-2)}\sigma^{(0)}_{ab} ), \ \ \text{for} \  d>3 \label{g1hd}
\end{eqnarray}
where $R_{\Sigma\ ab}$ is the curvature with respect to $\sigma^{(0)}_{ab}$. 

We use the background-field method developed in \cite{Miao:2013nfa} to investigate the holographic Weyl anomaly. This method is quite useful for the study of Weyl anomaly \cite{Miao:2013nfa}, correlation functions \cite{Sen:2014nfa}, entanglement/R\'enyi entropy \cite{Miao:2015iba,Miao:2015dua,Chu:2016tps} of higher derivative gravity.  Expanding $I_W$ eq.(\ref{IW}) in terms of background curvature $\bar{\mathcal{R}}$ defined below
\begin{eqnarray}\label{background curvature1}
&&\mathcal{R}=\bar{\mathcal{R}}-d(d-1)\text{sech}^2(\rho),\\ \label{background curvature2}
&& \mathcal{R}_{ij}=\bar{\mathcal{R}}_{ij}-(d-1)\text{sech}^2(\rho)h_{ij},\\ \label{background curvature3}
&&\mathcal{R}_{ijkl}=\bar{\mathcal{R}}_{ijkl}-\text{sech}^2(\rho)(h_{ik}h_{jl}-h_{il}h_{jk}),
\end{eqnarray}
we get 
 \begin{eqnarray}\label{IWanomaly}
I_{W}=I_{\text{CFT}}+I_{\text{W Ein}}+\frac{1}{16 \pi G_N}\int_Q \sqrt{|h|} \Big[\frac{\left(\bar{\mathcal{R}}^{ij}\bar{\mathcal{R}}_{ij}-\frac{d}{4(d-1)}\bar{\mathcal{R}}^2\right)}{(d-4)(d-2)^2}+\mathcal{L}_{\mathcal{R}3}(\bar{\mathcal{R}})+O(\bar{\mathcal{R}}^4)\Big],
 \end{eqnarray}
where $I_{\text{W Ein}}$ eq.(\ref{IWEinstein1}) is the action of Einstein gravity with an effective Newton's constant and AdS radius
 \begin{eqnarray}\label{Newton's constant}
\frac{1}{G_{N \text{ eff}}}=\frac{1}{ G_N}\int_{0}^{\rho}\frac{\cosh^{d-2}(r)}{\cosh^{d-2}(\rho)} dr,\ \ \ \ L_{\text{eff}}=\cosh(\rho).
\end{eqnarray}
Recall that for the class of solutions eq.(\ref{bulkmetric}), we have $I_{\text{CFT}}=0$ and $\bar{\mathcal{R}}_{ij}=\bar{\mathcal{R}}=\mathcal{L}_{\mathcal{R}3}(\bar{\mathcal{R}})=O(\bar{\mathcal{R}}^4)=0$. As a result, we get $I_{W}=I_{\text{W Ein}}$ for solutions eq.(\ref{bulkmetric}), which agrees with the discussions of sect.2.1.  
It should be stressed that, in general, $I_{W}$ and $I_{\text{W Ein}}$ are different.

 We are interested of only the Weyl anomaly for 2d, 4d and 6d CFTs ($d=3,5,7$). 
According to \cite{Miao:2013nfa}, we have for general solutions
\begin{eqnarray}\label{orders}
&&\sqrt{|h|}\left(\bar{\mathcal{R}}^{ij}\bar{\mathcal{R}}_{ij}-\frac{d}{4(d-1)}\bar{\mathcal{R}}^2\right) \sim O(z^{8-d}),\\
&&\sqrt{|h|} \mathcal{L}_{\mathcal{R}3}(\bar{\mathcal{R}}) \sim \sqrt{|h|}O(\bar{\mathcal{R}}^4) \sim O(z^{8-d}).
\end{eqnarray} 
Thus, the $O(\bar{\mathcal{R}}^2), O(\bar{\mathcal{R}}^3)$ and $O(\bar{\mathcal{R}}^4)$ terms of eq.(\ref{IWanomaly}) are irrelevant to the UV logarithmic divergent term of the effective action for $d< 9$.  Recall that $I_{\text{CFT}}$ is irrelevant to the UV logarithmic divergent term too.  Thus only the Einstein action $I_{\text{W Ein}}$ of eq.(\ref{IWanomaly}) contributes to the holographic Weyl anomaly. As a result, the holographic Weyl anomaly of wedge holography
 $\text{AdSW}_{d+1}/\text{CFT}_{d-1}$ is exactly the same as that of $\text{AdS}_{d}/\text{CFT}_{d-1}$ with Einstein gravity, provided that the Newton's constant and AdS radius are given by the effective ones eq.(\ref{Newton's constant}).

For the convenience of readers, we list the holographic Weyl anomaly for  2d, 4d and 6d CFTs below
\begin{eqnarray}\label{Weylanomaly2d}
&&\mathcal{A}_{2d}=\int_{\Sigma} dx^{2}
\sqrt{|\sigma|} \frac{c_{2d}}{24 \pi} R_{\Sigma},\\
&&\mathcal{A}_{4d}=\int_{\Sigma} dx^{4}
\sqrt{|\sigma|} [\frac{c}{16\pi^2} C_{\Sigma}^{ijkl}C_{\Sigma\ ijkl}-\frac{a}{16\pi^2}(R_{\Sigma}{}^{ijkl}R_{\Sigma}{}_{ijkl}-4R_{\Sigma}{}^{ij}R_{\Sigma}{}_{ij}+R_{\Sigma}^2)  ] ,\label{Weylanomaly4d}\\
&&\mathcal{A}_{6d}=\int_{\Sigma} dx^{6} \sqrt{|\sigma|} [ \sum_{n=1}^3 B_n I_n  +2 A\  E_{6} ]\label{Weylanomaly6d},
\end{eqnarray}
where $C_{\Sigma\ ijkl}$ is the Weyl tensor on $\Sigma$, $I_n$ are the Weyl invariant terms constructed from curvatures and their covariant derivatives,
\begin{eqnarray}\label{I1I2}
&&I_1=C_{\Sigma\ kijl}C^{\Sigma\ imnj}C_{\Sigma\ m\ \ n}^{\ \ \ \ kl},\ \ I_2=C_{\Sigma\ ij}^{\ \ \ \ kl}C_{\Sigma\ kl}^{\ \ \ \ mn}C_{\Sigma\ mn}^{\ \ \ \ \ \ ij},\\
&&I_3=C_{\Sigma\ iklm}(\Box_{\Sigma} \delta^i_j+4R_{\Sigma}{}^i_j-\frac{6}{5}R_{\Sigma}\delta^i_j)C^{\Sigma\ jklm},
\end{eqnarray}
 and $E_{2p}$ is the Euler density defined by
\begin{eqnarray}\label{E2p}
E_{2p}(R_{\Sigma})=\frac{1}{(8\pi)^p\Gamma(p+1)} \delta^{i_1 i_2...i_{2p-1}i_{2p}}_{j_1 j_2...j_{2p-1}j_{2p}} R_{\Sigma}{}^{j_1 j_2}_{\ \ \ \ i_1i_2}... R_{\Sigma}{}^{j_{2p-1} j_{2p}}_{\ \ \ \ \ \ \ \  i_{2p-1}i_{2p}}. 
\end{eqnarray}
For wedge holography, the central charges of above Weyl anomaly are given by 
\begin{eqnarray}\label{charge2d}
&& c_{2d}=\frac{3}{2 G_N}\int_0^{\rho} \cosh(r)dr=\frac{3}{2 G_N} \sinh(\rho),\\ \label{charge4d}
&& a=c=\frac{\pi}{8 G_N}\int_0^{\rho} \cosh^{3}(r)dr=\frac{\pi}{96 G_N}  (9 \sinh (\rho )+\sinh (3 \rho )) ,\\
&& A=\frac{\pi^2}{ 16 G_N}\int_{0}^{\rho}\cosh^{4}(r)dr, \  B_1=-\frac{1}{256 \pi G_N}\int_{0}^{\rho}\cosh^{4}(r)dr\label{charge6d1},\\
&& B_2=-\frac{1}{1024 \pi G_N }\int_{0}^{\rho}\cosh^{4}(r)dr, \ B_3=\frac{1}{3072\pi G_N }\int_{0}^{\rho}\cosh^{4}(r)dr.\label{charge6d2}
\end{eqnarray}
Recall that we focus on half of the wedge space $0\le r \le \rho$ in above discussions. For the whole  wedge space, we should double the above central charges.  Note also that although the metrics on the two branes $Q_1$ and $Q_2$ are different generally, the induced metrics $\sigma_{ab}$ on the corner $\Sigma$ are  the same. 

As a summary, we obtain the holographic Weyl anomaly for wedge holography with general solutions in this section. Remarkably, although the effective action on the brane are infinite towers of higher derivative gravity, the holographic Weyl anomaly is exactly the same as that of Einstein gravity. It should be mentioned that the holographic Weyl anomaly for 2d and 4d CFTs eqs.(\ref{Weylanomaly2d},\ref{Weylanomaly4d}) have been derived in \cite{Miao:2020oey} for the special class of solution eq.(\ref{bulkmetric}). Here we re-derive the same holographic Weyl anomaly for the most general solutions in wedge holography. Since Weyl anomaly is independent of the states of CFTs \footnote{CFTs in vacuum state and thermal state have the same central charges and thus the same Weyl anomaly.}, the holographic Weyl anomaly should be irrelevant to the bulk solutions. As we have shown in this section,  this is indeed the case. This self-consistency can be regarded as a test of wedge holography and the effective action eq.(\ref{IW}).

\section{Vector on the brane}

In this section, we study the mass spectrum and the perturbative effective action of vectors on the end-of-the-world brane. The case of gravity is similar.

\subsection{Mass spectrum}

\begin{figure}[t]
\centering
\includegraphics[width=8cm]{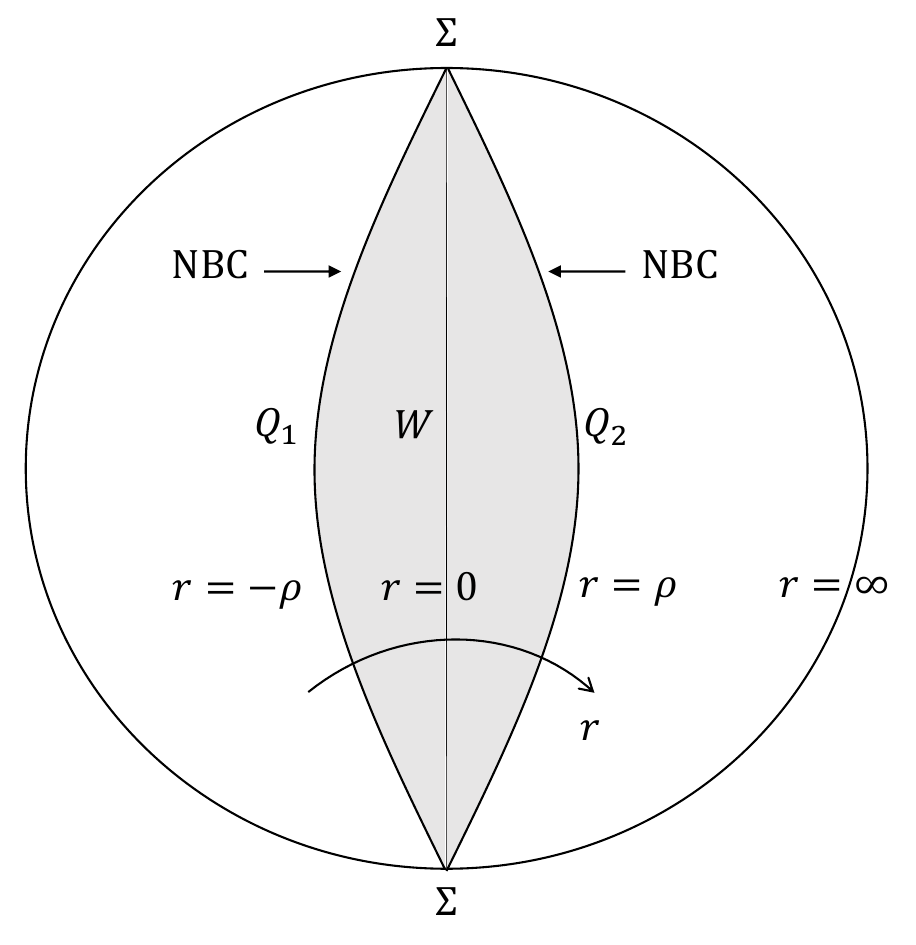}
\caption{Geometry of wedge holography associated with the coordinate system of eqs.(\ref{bulkmetric},\ref{probeQ}).}
\label{wedgegeometry}
\end{figure}

Let us first discuss the mass spectrum. For simplicity, we focus on the probe limit, where the bulk metric and the embedding function of $Q$ are given by
 \begin{eqnarray}\label{bulkmetric}
&&ds^2=dr^2+\cosh^2 (r) \bar{h}^{(0)}_{ij}(y) dy^i dy^j,\\
&& Q: \ r=\pm \rho. \label{probeQ}
\end{eqnarray}
See Fig.\ref{wedgegeometry} for the geometry, where the branes are located at $r=\pm \rho$, and $\bar{h}^{(0)}_{ij}(y)$ is the AdS metric with the unit radius $L=1$ on the branes.  We take the following ansatz of Maxwell's fields in the bulk
 \begin{eqnarray}\label{vector}
\mathcal{A}_r=0,  \ \mathcal{A}_i= S(r) A_i(y). 
\end{eqnarray}
Substituting eq.(\ref{vector}) into Maxwell's equations $\nabla_{\mu}\mathcal{F^{\mu\nu}}=0$ and separating variables, we obtain
 \begin{eqnarray}\label{EOMvector}
&& \bar{D}_i F^{ij}-m^2_v A^j=0,\\
&& \cosh^2(r) S''(r)+(d-2) \sinh (r) \cosh (r) S'(r) + m_v^2 S(r)=0, \label{EOMmassiveS}
\end{eqnarray}
where $F=dA$, $\bar{D}_i$ are the covariant derivatives with respect to $\bar{h}^{(0)}_{ij}(y)$, and $m_v$ denotes the mass of vectors. Solving eq.(\ref{EOMmassiveS}), we get
 \begin{eqnarray}\label{massiveSsolution}
S(r)=\text{sech}^{\frac{d-2}{2}}(r) \left(c_1 P_{\lambda_v}^{\frac{d-2}{2}}(\tanh r)+c_2 Q_{\lambda_v}^{\frac{d-2}{2}}(\tanh r)\right),
\end{eqnarray}
where $P_{\lambda_v}^{\frac{d-2}{2}}$ and $ Q_{\lambda_v}^{\frac{d-2}{2}}$ are the Legendre polynomials, $c_1$ and $c_2$ are integral constants and $\lambda_v$ is given by
 \begin{eqnarray}\label{nuvector}
\lambda_v=\frac{1}{2} \left(\sqrt{(d-3)^2+4  m_v^2}-1\right).
\end{eqnarray}
From eq.(\ref{nuvector}), we derive the expected  Breitenlohner-Freedman (BF) bound of massive vectors in 
$\text{AdS}_d$
 \begin{eqnarray}\label{boundvectormass}
m_v^2\ge-(\frac{d-3}{2})^2.
\end{eqnarray}

One can impose either the absolute BC or the relative BC for vectors on the 
end-of-the-world branes
\begin{equation}\label{BCVector}
\begin{split}
&\text{absolute BC} : \mathcal{F}_{ni}|_{Q}=0,\\
&\text{relative BC} : {}^*\mathcal{F}_{ni}|_{Q}=0,
\end{split}
\end{equation}
where $n$ denotes the normal direction, $i$ denotes the tangent direction, $\mathcal{F}=d\mathcal{A}$ is the bulk field strength and ${}^*\mathcal{F}$ is the Hodge dual of $\mathcal{F}$.  
For our ansatz eqs.(\ref{bulkmetric},\ref{probeQ},\ref{vector}), the absolute BC and relative BC become Neumann BC (NBC) and Dirichlet BC (DBC), respectively
\begin{eqnarray}\label{BCVector1NBC}
&&\text{NBC} : S'(\pm \rho)=0,\\
&&\text{DBC} : S(\pm \rho)=0. \label{BCVector1DBC}
\end{eqnarray}
The dynamical field on the brane is the induced vector and its conjugate momentum for NBC and DBC, respectively
\be
\begin{cases} 
 \d \mathcal{A}_i= \ S(\pm \rho) A_i,
  & \mbox{for absolute BC/NBC},\\
  \d \mathcal{F}_{ni}= \pm S'(\pm \rho) A_i, & \mbox{for relative BC/DBC}.
\end{cases}
\ee

 Now we are ready to study the mass spectrum of vectors on the end-of-the-world branes.  Let us first study the case of NBC.  For the following purpose, we denote $S(r)$ by $S(r)=c_1 \bar{P}(r) + c_2 \bar{Q}(r)$ where $\bar{P}(r)=\text{sech}^{\frac{d-2}{2}}(r)P_{\lambda_v}^{\frac{d-2}{2}}(\tanh r)$ and $\bar{Q}(r)=\text{sech}^{\frac{d-2}{2}}(r)Q_{\lambda_v}^{\frac{d-2}{2}}(\tanh r)$. 
Imposing NBC eq.(\ref{BCVector1NBC}), we get 
 \begin{eqnarray}\label{matrix}
M \cdot c=\left( \begin{matrix}
\bar{P}'(\rho)&\bar{Q}'(\rho)\\
\bar{P}'(-\rho)&\bar{Q}'(-\rho)
\end{matrix} \right) \cdot \left( \begin{matrix}
c_1\\
c_2
\end{matrix} \right)=0.
\end{eqnarray}
To have non-trivial solutions of $c_1$ and $c_2$, we must have 
\begin{eqnarray}\label{detM}
|M|=\bar{P}'(\rho)\bar{Q}'(-\rho)-\bar{Q}'(\rho)\bar{P}'(-\rho)=0,
\end{eqnarray}
 which gives a constraint for the mass of the vector
\begin{eqnarray}\label{constraintmassvectorcase1}
\text{NBC}:\ m_v^2 \ \text{sech}^{d}(\rho )  \left(P_{\lambda _v}^{\frac{d}{2}-2}(x) Q_{\lambda _v}^{\frac{d}{2}-2}(-x)-P_{\lambda _v}^{\frac{d}{2}-2}(-x) Q_{\lambda _v}^{\frac{d}{2}-2}(x)\right)=0 ,
\end{eqnarray}
where $x=\tanh(\rho)$ and $\lambda_v$ is a function of the mass  eq.(\ref{nuvector}).  

It is clear that the massless mode with $m_v=0$ is a solution to eq.(\ref{constraintmassvectorcase1}) for wedge holography with NBC.  
On the other hand, there is no massless mode for the usual brane world theory \footnote{For simplicity, we focus on one side of the brane in brane world theory, where the junction condition is equivalent to NBC.} and AdS/BCFT defined in the region $-\rho\le r\le \infty$ with NBC on the brane $S'(-\rho)=0$ and DBC $S(\infty)=0$ on the AdS boundary.  Let us explain more on this point.  Solving eq.(\ref{EOMmassiveS}) with $m_v=0$, we get
\begin{eqnarray}\label{Sm0}
S(r)=c_1+ c_2 \int_0^r \text{sech}^{d-2}(s) ds.
\end{eqnarray}
Imposing NBC $S'(\pm \rho)=0$ for wedge holography, we derive
\begin{eqnarray}\label{Sm01}
S(r)=c_1,
\end{eqnarray}
which yields non-zero induced vector $\mathcal{A}_i= S(\pm \rho) A_i=c_1 A_i$ on the brane.  On the other hand,  the BCs  $S'(-\rho)=0$ and $S(\infty)=0$ for brane world theory and AdS/BCFT yield $S(r)=0$. As a result, there is no massless mode in the usual  brane world holography and AdS/BCFT.  
We take the existence of massless modes as an advantage of the wedge holography with NBC, since there are Maxwell's fields on the end-of-the-world branes. 
Furthermore, the massless mode can be normalized since wedge holography is defined in a finite region $-\rho\le r \le \rho$. 
However, since $S(r)$ eq.(\ref{Sm01}) is a constant in the bulk, the massless vector is not located perfectly on the brane. By perfect localization, we mean that the wave function $S(r)$ peaks on the brane only and decays when it goes far from the brane.  
The massless mode is a critical case: the wave function $S(r)$ neither increases nor decreases but keeps a constant when it goes far from the brane.  The localization looks better in the coordinate that the EOM eq.(\ref{EOMvector}) takes the form of schrödinger equation. See  sect. 3.1 for more discussions.

Similar to \cite{Chu:2021mvq}, there is no solution to \eq{constraintmassvectorcase1} for $m^2 < -(d-3)^2/4$ outside the  Breitenlohner-Freedman (BF) bound. Naively, the negative value of $m^2$ within the BF bound
\begin{equation}\label{neg-m2}
  m_{v}^2 =\begin{cases}
  -\frac{1}{4}((d-3)^2-1),&\ \text{even $d$},\\
  -\frac{1}{4}(d-3)^2,&\ \text{odd $d$},
\end{cases}
\end{equation}
 is a solution to eq.(\ref{constraintmassvectorcase1}). However, according to \cite{Chu:2021mvq}, for these special values, \eq{massiveSsolution} no longer gives the general solution since
the Legendre functions $P^n_0(x),     Q^{n+\frac{1}{2}}_{-\frac{1}{2}}(x)$ vanish
identically for integer $n$. In this case,
the general solution is given by
\begin{equation}\label{S-speca}
  S(r)=\begin{cases}
  \text{sech}^{\frac{d-2}{2}}(r) \left(c_1 e^{\frac{d-2}{2}r} + c_2 e^{-\frac{d-2}{2}r}
\right),&\ \text{even $d$},\\
 \text{sech}^{\frac{d-2}{2}}(r) \left(
c_1 P^{\frac{d-2}{2}}_{-\frac{1}{2}}(\tanh r) + c_2 P^{-\frac{d-2}{2}}_{-\frac{1}{2}}(\tanh r)
\right),&\ \text{odd $d$}.
\end{cases}
\end{equation}
For $\rho>0$, the above solutions cannot satisfy NBC $S'(\pm \rho)=0$, so we should rule out \eq{neg-m2} from the spectrum. As a result,  we have $m_v^2\ge 0$ for the mass spectrum of vectors with NBC.  

\begin{figure}[t]
\centering
\includegraphics[width=7.5cm]{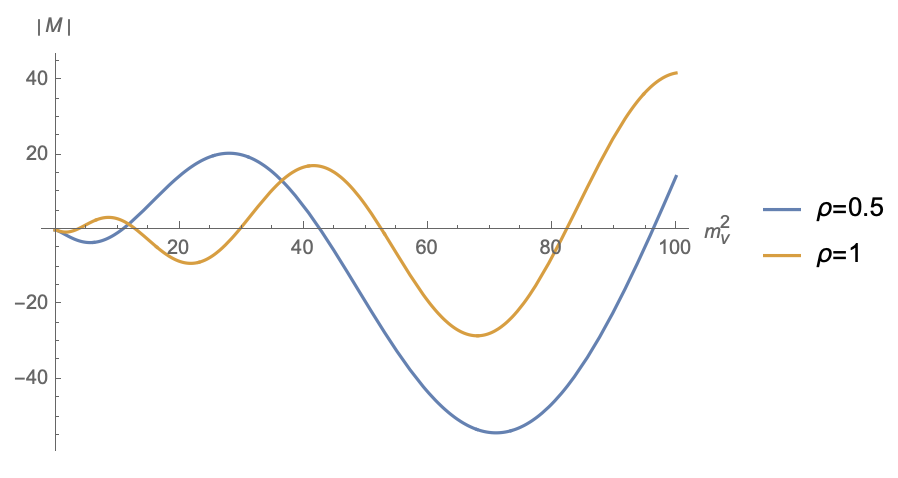}
\includegraphics[width=7.5cm]{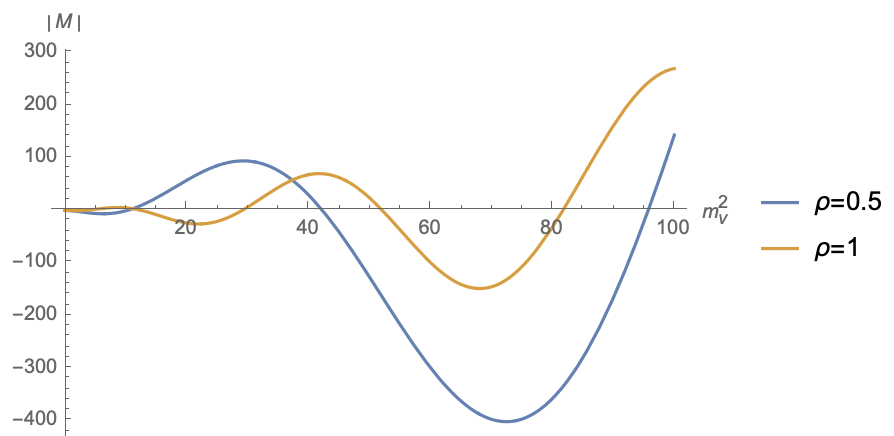}
\caption{Mass spectrum for vectors with NBC. The left figure is for $d=3$, and the right figure is for $d=4$,  $|M|$ is given by eq.(\ref{constraintmassvectorcase1}) and the roots of $|M|=0$ denote the mass squares. The massless mode is always a solution to NBC. Besides, the larger the tension $T=(d-1)\tanh(\rho)$ is, the continuous the mass spectrum is. }
\label{vectormass3d4d}
\end{figure}

Let us draw some figures to illustrate the spectrum.  See Fig.\ref{vectormass3d4d}, where the intersections of the curves and $m_v^2$-axis denote the mass squares.  As shown in Fig.\ref{vectormass3d4d}, the massless mode is indeed a solution.  Besides, the larger the tension $T=(d-1)\tanh(\rho)$ is, the continuous the mass spectrum is.

To get more understandings of the spectrum, let us study two limits.  For the large $\rho\to \infty$ ($x\to 1$), eq.(\ref{constraintmassvectorcase1}) can be approximately by
\begin{equation}\label{constraintonMlarge}
0\sim \text{sech}^{\frac{d}{2}+2}(\rho)  \begin{cases}
 \sin(\lambda_v \pi) \,&\ \text{even d} ,\\
 \cos(\lambda_v  \pi),&\ \text{odd d},
\end{cases}
\end{equation}
which has the roots
\begin{equation}\label{masslargerhovector}
m_v^2 \approx 
k (k+d-3), \, \ \ \ \ \text{for large } \rho,
\end{equation}
where $k\ge 0$ are integers.   
In the contrary limit of small $\rho$, $-\rho\le r \le \rho$ is also small. Then eq.(\ref{EOMmassiveS}) becomes approximately by
 \begin{eqnarray}
S''(r) + m_v^2 S(r)=0, \label{EOMMBCmassiveHsmallrho}
\end{eqnarray}
which can be solved as
 \begin{eqnarray}\label{massiveHsolutionsmallrho}
S(r)=c_1 \cos(|m_v| r)+ c_2 \sin(|m_v| r).
\end{eqnarray}
Imposing NBC eq.(\ref{BCVector1NBC}) and following the above approach, we get
 \begin{eqnarray}\label{masssmallrhoeq}
m_v^2 \sin(2|m_v| \rho)=0,
\end{eqnarray}
which yields the mass spectrum
\begin{equation}\label{masssmallrhovector}
m_v^2 \approx
\frac{k^2\pi^2}{4\rho^2}, \, \ \ \ \ \text{for small } \rho.
\end{equation}

As a summary, the mass spectrum of vectors with NBC is non-negative, i.e., $m_v^2\ge 0$. In particular, it contains a massless mode, which is quite different from the brane world holography and AdS/BCFT. 
For the large $\rho$, the mass spectrum is given by eq.(\ref{masslargerhovector}). While for small $\rho$, the mass spectrum is approximately by eq.(\ref{masssmallrhovector}), which is independent of the dimensions of spacetime. 

\begin{figure}[t]
\centering
\includegraphics[width=7.5cm]{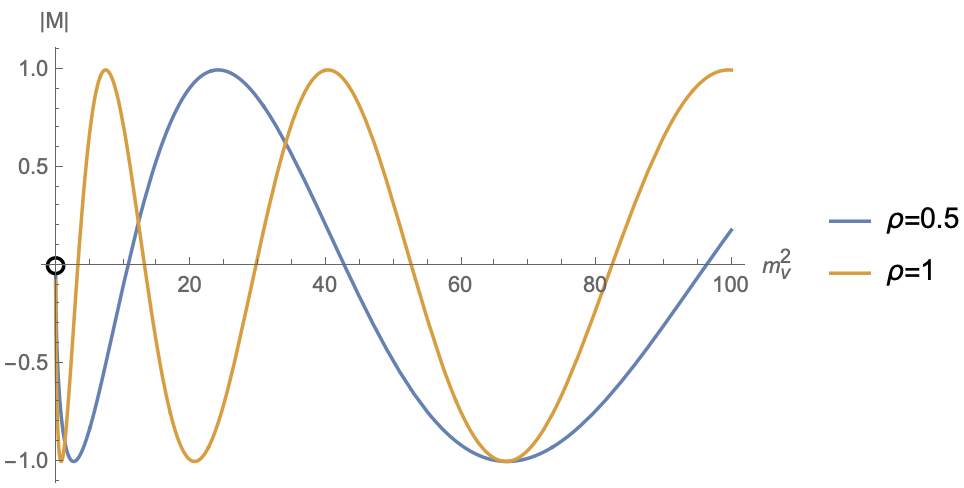}
\includegraphics[width=7.5cm]{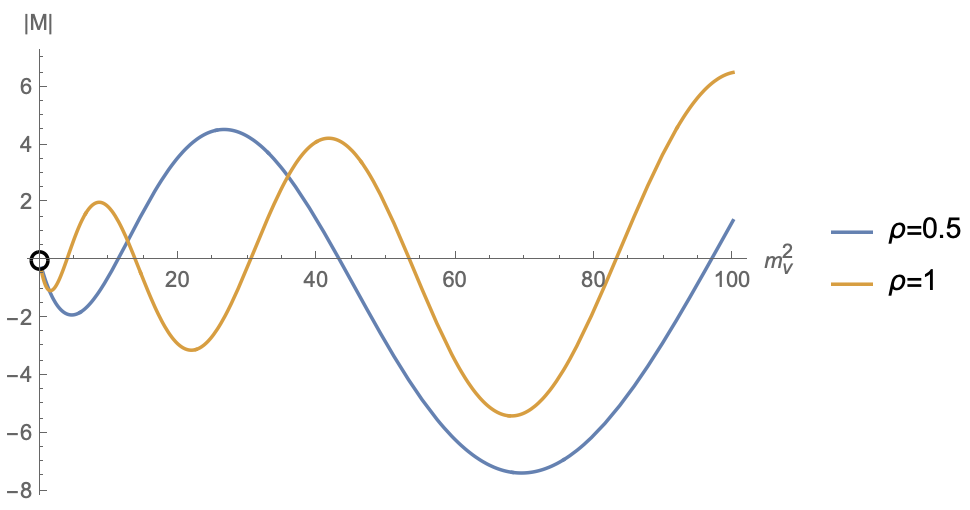}
\caption{Mass spectrum for vectors with DBC. The left figure is for $d=3$, and the right figure is for $d=4$,  $|M|$ is given by eq.(\ref{constraintmassvectorcase1DBC}) and the roots of $|M|=0$ denote the mass squares. 
The small circle at origin means that the massless mode is removed from the mass spectrum for DBC. The larger the tension $T=(d-1)\tanh(\rho)$ is, the continuous the mass spectrum is. }
\label{vectormass3d4dDBC}
\end{figure}

Let us go on to study the mass spectrum of DBC eq.(\ref{BCVector1DBC}). Since the approach is quite similar to that of NBC  eq.(\ref{BCVector1NBC}), we do not repeat the calculations but just list main results below. The constraint of the mass spectrum with DBC is given by
\begin{eqnarray}\label{constraintmassvectorcase1DBC}
\text{DBC}:\  \text{sech}^{d-2}(\rho )  \left(P_{\lambda _v}^{\frac{d}{2}-1}(x) Q_{\lambda _v}^{\frac{d}{2}-1}(-x)-P_{\lambda _v}^{\frac{d}{2}-1}(-x) Q_{\lambda _v}^{\frac{d}{2}-1}(x)\right)=0,
\end{eqnarray}
where $x=\tanh(\rho)$. The mass spectrum of vectors with DBC includes only positive $m_v^2> 0$. Unlike NBC, the massless mode is no longer a solution to DBC.  One can easily check that the solution eq.(\ref{Sm0}) with $m_v=0$ cannot satisfy the DBC $S(\pm \rho)=0$ unless $S(r)=0$. In the large and small $\rho$ limit, except a massless mode, the mass spectrum of DBC is exactly the same as that of NBC
\begin{equation}\label{massvectorDBClargeandsmall}
\text{DBC}:\  m_{v}^2  \approx \begin{cases}
 l (l+d-3),&\ \text{for large $\rho$},\\
\frac{l^2\pi^2}{4\rho^2},&\ \text{for small $\rho$},
\end{cases}
\end{equation}
where $l>0$ is a positive integer.   
See Fig. \ref{vectormass3d4dDBC} for the mass spectrum of vectors with DBC. 

\subsection{Perturbative action}

Let us go on to study the perturbative action of vectors on the end-of-the-world branes.  We expand the bulk vectors in infinite powers of Kaluza-Klein (KK) modes 
 \begin{eqnarray}\label{vector-power}
\mathcal{A}_r=0,  \ \mathcal{A}_i= \sum_{m_v}S^{(m_v)}(r) A^{(m_v)}_i(y),
\end{eqnarray}
where $\sum_{m_v}$ denotes the sum over the mass spectrum, $A^{(m_v)}_i(y)$ and $S^{(m_v)}(r)$ obey EOM eqs.(\ref{EOMvector},\ref{EOMmassiveS}) and the orthogonal condition
 \begin{eqnarray}\label{orthogonal-vector}
\int_{-\rho}^{\rho}\cosh(r)^{d-4}S^{(m_v)}(r) S^{(m'_v)}(r) dr=\delta^{m_v, m'_v}. 
\end{eqnarray}
Substituting eq.(\ref{vector-power}) into the bulk action
 \begin{eqnarray}\label{bulkaction-vector}
I=-\frac{1}{4}\int_N d^{d+1}x\sqrt{|g|} \mathcal{F}_{\mu\nu}  \mathcal{F}^{\mu\nu},
\end{eqnarray}
we get
 \begin{eqnarray}\label{bulkaction-vector1}
&&I=-\frac{1}{4}\sum_{m_v,m'_v}\int_{-\rho}^{\rho}dr \int_Q d^{d}y\sqrt{|\bar{h}^{(0)}|} \Big{[} 2 \cosh(r)^{d-2} S'^{(m_v)}(r)S'^{(m'_v)}(r)A^{(m_v)}_i A^{(m'_v)i}\nonumber\\
&&\ \ \ \ \ \ \ \ \ \ \ \ \ \ \ \ \ \ \ \ \ \ \ \ \ \ \ \ \ \ \ \ \ + \cosh(r)^{d-4} S^{(m_v)}S^{(m'_v)}F^{(m_v)}_{ij}F^{(m'_v)ij}\Big{]},
\end{eqnarray}
where $A^i$ and $F^{ij}$ are raised by $\bar{h}^{(0)ij}$. Integrating by parts and imposing either NBC $S'(\pm \rho)=0$ or DBC $S(\pm \rho)=0$, we derive
 \begin{eqnarray}\label{bulkaction-vector2}
&&I=-\frac{1}{4}\sum_{m_v,m'_v}\int_{-\rho}^{\rho}dr \int_Q d^{d}y\sqrt{|\bar{h}^{(0)}|} \Big{[} -\frac{d}{dr}\left(2 \cosh(r)^{d-2} S'^{(m_v)}(r) \right)S^{(m'_v)}(r)A^{(m_v)}_i A^{(m'_v)i}\nonumber\\
&&\ \ \ \ \ \ \ \ \ \ \ \ \ \ \ \ \ \ \ \ \ \ \ \ \ \ \ \ \ \ \ \ \ + \cosh(r)^{d-4} S^{(m_v)}S^{(m'_v)}F^{(m_v)}_{ij}F^{(m'_v)ij}\Big{]}.
\end{eqnarray}
By using EOM eq.(\ref{EOMmassiveS}) and orthogonal condition eq.(\ref{orthogonal-vector}), the above action can be simplified as
 \begin{eqnarray}\label{bulkaction-vector3}
&&I=-\frac{1}{4}\sum_{m_v}\int_{-\rho}^{\rho}dr \cosh(r)^{d-4}S^{(m_v)}(r)^2  \int_Q d^{d}y\sqrt{|\bar{h}^{(0)}|} \Big{[} 2m_v^2A^{(m_v)}_i A^{(m_v)i}+F^{(m_v)}_{ij}F^{(m_v)ij}\Big{]}\nonumber\\
&&\ \ \ =-\frac{1}{4}\sum_{m_v}\int_Q d^{d}y\sqrt{|\bar{h}^{(0)}|} \Big{[} 2m_v^2A^{(m_v)}_i A^{(m_v)i}+F^{(m_v)}_{ij}F^{(m_v)ij}\Big{]},
\end{eqnarray}
which is the correct action of massive vectors.

\section{Gravity on the brane}

In this section, we investigate the mass spectrum and the perturbative effective action of  gravitons on the end-of-the-world branes. Interestingly, the mass spectrum of gravitons in $d$ dimensions is the same as that of vectors in $(d+2)$ dimensions. In particular, there is a massless mode when one imposes NBC on both branes.

\subsection{Mass spectrum}

Let us first discuss the mass spectrum of gravitons on the end-of-the-world branes. 
We choose the following ansatz of the perturbation metric and the embedding function of $Q$
 \begin{eqnarray}\label{perturbationmetric}
&&ds^2=dr^2+\cosh^2 (r) \left( \bar{h}^{(0)}_{ij}(y) + \epsilon H(r) \bar{h}^{(1)}_{ij}(y)  \right)dy^i dy^j+O(\epsilon^2),\\
&& Q: \ r=\pm \rho+O(\epsilon^2), \label{perturbationQ}
\end{eqnarray}
where $\bar{h}^{(0)}_{ij}(y)$ is the AdS metric with a unit radius and $\bar{h}^{(1)}_{ij}(y)$ denotes the perturbation. 
In terms of bulk metric perturbations,  we have
 \begin{eqnarray}\label{bulkmetricperturbations}
\delta g_{r\mu}=0,\ \delta g_{ij}=\cosh^2 (r)  H(r) \bar{h}^{(1)}_{ij}(y).
\end{eqnarray}
Imposing the transverse traceless gauge 
 \begin{eqnarray}\label{gijgauge}
\nabla^{\mu} \delta g_{\mu\nu}=0,\ \ \  g^{\mu\nu}\delta g_{\mu\nu}=0,
\end{eqnarray}
we get
 \begin{eqnarray}\label{hij1gauge}
\bar{D}^i \bar{h}^{(1)}_{ij}=0,\ \ \  \bar{h}^{(0)ij}\bar{h}^{(1)}_{ij}=0,
\end{eqnarray}
where $\nabla_{\mu}$ and $\bar{D}_i$ are the covariant derivatives with respect to $g_{\mu\nu}$ and $\bar{h}^{(0)}_{ij}$, respectively. 
Note that the gauge $\bar{h}^{(0)ij}\bar{h}^{(1)}_{ij}=0$ eliminates the scalar mode in the metric fluctuation. Since there is only a scalar degree of freedom for the metric in two dimensions,  the gauge eq.(\ref{hij1gauge}) removes all the degree of freedom for the 2-dimensional metric. As a result, the discussions of this section works only for $d\ge 3$.

One can impose either NBC \cite{Takayanagi:2011zk}, DBC \cite{Miao:2018qkc} or CBC \cite{Chu:2021mvq} on the end-of-the-world branes
\begin{eqnarray}\label{NBCH}
&&\text{NBC}:\ \ \ \ \ \ \ \ H'(\pm \rho)=0, \\
&&\text{DBC/CBC}: H(\pm \rho)=0, \label{DBCH}
\end{eqnarray}
where CBC specifies the conformal geometry of the boundary and the trace of the extrinsic curvature. Note that DBC and CBC are the same at the linear order of perturbations, they are different at higher orders generally \cite{Chu:2021mvq}. 
For DBC/CBC, we have the freedom to choose the induced metric on the brane. For simplicity, we choose the same induced metric, the same background metric eq.(\ref{bulkmetric}) in the bulk and the same location of branes eq.(\ref{perturbationQ}) as those of NBC.  Remarkably, the mass spectrum is independent of the choice of the induced metric $ \bar{h}^{(0)}_{ij}(y)$, as long as it satisfies Einstein equations eq.(\ref{EinsteinEQonbrane}) on the brane. That is because, for any $ \bar{h}^{(0)}_{ij}(y)$ obeying eq.(\ref{EinsteinEQonbrane}), the boundary conditions eqs.(\ref{NBCH},\ref{DBCH}) and the equation of motion of $H(r)$ eq.(\ref{EOMMBCmassiveH}) are the same. As a result, the mass spectrum determined by the boundary condition and the equation of motion of $H(r)$ is independent of the choice of the induced metric $ \bar{h}^{(0)}_{ij}(y)$, as long as it obeys eq.(\ref{EinsteinEQonbrane}).

In the gauge eq.(\ref{gijgauge}), Einstein equations become
 \begin{eqnarray}\label{EOMgABgauge}
 \left(\nabla_{\alpha} \nabla^{\alpha}+2\right) \delta g_{\mu\nu}=0.
\end{eqnarray}
Substituting eq.(\ref{bulkmetricperturbations}) together with eq.(\ref{hij1gauge}) into eq.(\ref{EOMgABgauge}) and separating variables, we obtain
 \begin{eqnarray}\label{EOMMBCmassivehij}
&& \left(\bar{D}_k \bar{D}^k+2-m_g^2\right)\bar{h}^{(1)}_{ij}(y)=0,\\
&& \cosh^2(r) H''(r)+d \sinh (r) \cosh (r) H'(r) + m_g^2 H(r)=0, \label{EOMMBCmassiveH}
\end{eqnarray}
where $m_g$ denotes the mass of gravitons.  
Note that we have assumed that $\bar{h}^{(0)}_{ij}(y)$ is an AdS metric in the above derivations. For general $\bar{h}^{(0)}_{ij}(y)$ obeying eq.(\ref{EinsteinEQonbrane}), eq.(\ref{EOMMBCmassiveH}) is unchanged but eq.(\ref{EOMMBCmassivehij}) becomes
 \begin{eqnarray}\label{EOMMBCmassivehijnonAdS}
&& \left(\bar{D}_k \bar{D}^k+2-m_g^2\right)\bar{h}^{(1)}_{ij}(y)+2C^{(0)}_{imjn}\bar{h}^{(1)}{}^{mn}(y)=0,
\end{eqnarray} 
where $C^{(0)}_{imjn}$ is the Weyl tensor defined by $\bar{h}^{(0)}_{ij}(y)$.

Solving eq.(\ref{EOMMBCmassiveH}), we get
 \begin{eqnarray}\label{massiveHsolution}
H(r)=\text{sech}^{\frac{d}{2}}(r) \left(c_1 P_{\lambda_g}^{\frac{d}{2}}(\tanh r)+c_2 Q_{\lambda_g}^{\frac{d}{2}}(\tanh r)\right),
\end{eqnarray} 
where $P_{\lambda_g}^{\frac{d}{2}}$ and $ Q_{\lambda_g}^{\frac{d}{2}}$ are the Legendre polynomials, $c_1$ and $c_2$ are integral constants and $\lambda_g$ is given by
 \begin{eqnarray}\label{aibia1}
\lambda_g=\frac{1}{2} \left(\sqrt{(d-1)^2+4  m_g^2}-1\right),
\end{eqnarray}
which yields the correct BH bound of massive gravity in 
$\text{AdS}_d$
 \begin{eqnarray}\label{BFgravity}
 m_g^2\ge -(\frac{d-1}{2})^2.
\end{eqnarray}
Recall that we have set the AdS radius $L=1$. 

Remarkably, the EOM of $H(r)$ eq.(\ref{EOMMBCmassiveH}) in $d$ dimensions is exactly the same as that of $S(r)$ eq.(\ref{EOMmassiveS}) in $(d+2)$ dimensions. Furthermore, $H(r)$ and $S(r)$ obey the same boundary conditions eqs.(\ref{BCVector1NBC},\ref{BCVector1DBC},\ref{NBCH},\ref{DBCH}).  As a result, the mass spectrum of gravitons in $d$ dimensions is the same as the spectrum of vectors in $(d+2)$ dimensions. Thus there is no need to repeat the calculations again. For the convenience of readers, we list the key characteristics of the gravitational spectrum below. 

{\bf 1.} The gravitational spectrum obeys the following constraint 
\begin{eqnarray}\label{constraintmassgravityNBC}
&&\text{NBC}:\ m_g^2 \ \text{sech}^{d+2}(\rho )  \left(P_{\lambda _g}^{\frac{d}{2}-1}(x) Q_{\lambda _g}^{\frac{d}{2}-1}(-x)-P_{\lambda _g}^{\frac{d}{2}-1}(-x) Q_{\lambda _g}^{\frac{d}{2}-1}(x)\right)=0,\\
&&\text{DBC/CBC}:\  \text{sech}^{d}(\rho )  \left(P_{\lambda _g}^{\frac{d}{2}}(x) Q_{\lambda _g}^{\frac{d}{2}}(-x)-P_{\lambda _g}^{\frac{d}{2}}(-x) Q_{\lambda _g}^{\frac{d}{2}}(x)\right)=0, \label{constraintmassgravityDBCCBC}
\end{eqnarray}
where $x=\tanh(\rho)$ and $\lambda_g=\lambda_v(d\to d+2)$ is given by eq.(\ref{aibia1}).

{\bf 2.}  We have $m_g^2\ge 0$ for NBC and  $m_g^2> 0$ for DBC/CBC. In particular, there is a massless mode when one imposes NBC on both branes.  This  is consistent with the result of \cite{Miao:2020oey}, which finds that, for one class of solutions to NBC, the effective gravity on the branes is Einstein gravity.  On the other hand, there is no massless mode in the brane-world holography \cite{Randall:1999ee,Randall:1999vf,Karch:2000ct} unless the brane tension approaches the critical value $T=(d-1)$.

\begin{figure}[t]
\centering
\includegraphics[width=7.5cm]{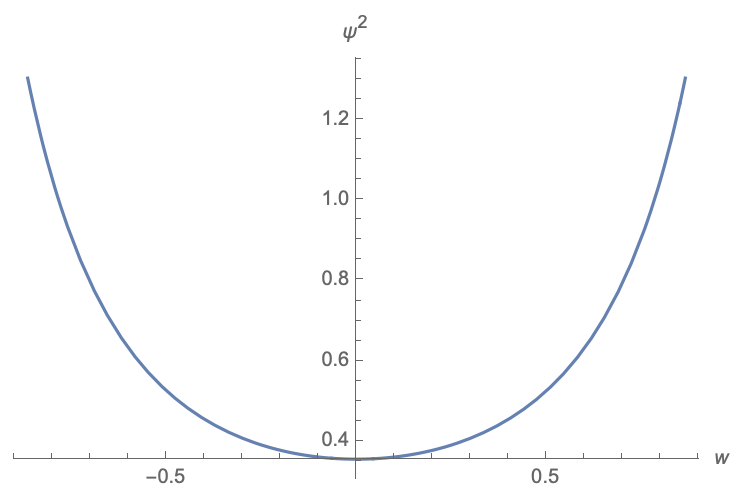}
\includegraphics[width=7.5cm]{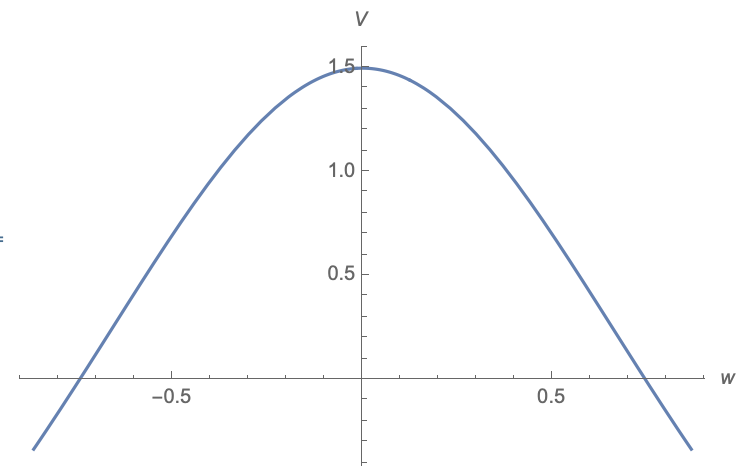}
\caption{(Left) wave function of massless graviton, which shows that the massless graviton is located on the branes at $w=\pm w_0$; (Right) ``volcano potential", where the negative delta function potential on the brane is ignored. We have $d=4$ and $w_0\approx0.87$ ($\rho=1$) for above figures.  }
\label{wavefunction}
\end{figure}

{\bf 3.} 
Let us make some comments on the localization for the massless mode on the brane.  By localization, we mean that the wave function peaks on the brane only and decays when it goes far from the brane.  
 In the geodesic coordinate $r$, the wave function $H(r)$ is a constant for the massless mode. Thus, it is not located perfectly on the brane. Instead, it is a critical case.  In the coordinate $w$ defined below, the localization looks better.  Performing
the transformations
  \begin{eqnarray}\label{transformationsNBC}
    dw=\frac{dr}{\cosh(r)}, \  \Psi(w)=\cosh^{\frac{d-1}{2}}(r) H(r),
 \end{eqnarray}
 we rewrite eq.(\ref{EOMMBCmassiveH}) into the form of schrödinger equation
  \begin{eqnarray}\label{EOMII}
-\Psi''(w)+ V(w)  \Psi(w)=m_g^2 \Psi(w),
  \end{eqnarray}
where the ``volcano potential" is given by
\begin{eqnarray}\label{VNBC}
  V(w)=\frac{1}{4} (d-1)
  \Big( (d+1) \sec ^2(w)-d+1\Big)-(d-1)\sinh (\rho )
 \delta(w\pm w_0),
 \end{eqnarray}
 where $w_0=\sin^{-1}(\text{tanh}(\rho))$ and the two branes are located at $w=\pm w_0$.  Due to the ``volcano potential", the massless mode tends to be located on the branes at low energies. Indeed, as shown in Fig. \ref{wavefunction}, the wave function $\Psi(w)$ peaks on the brane only and decays when it goes far from the brane. However, $\Psi(w)$ does not vanish in the middle of the wedge space $w=0$. Thus, the massless mode is not perfectly located on the brane. 
 
{\bf 4.} In the large and small $\rho$ limit, except a massless mode, the mass spectrum of NBC is exactly the same as that of DBC/CBC
\begin{equation}\label{massgravitysmallandlarge}
\text{NBC/DBC/CBC}:\  m_{g}^2  \approx \begin{cases}
 k (k+d-1),&\ \text{for large $\rho$},\\
\frac{k^2\pi^2}{4\rho^2},&\ \text{for small $\rho$},
\end{cases}
\end{equation}
where $k$ is an integer, $k\ge 0$ for NBC and $k> 0$ for DBC/CBC. 
 Recall that $m^2_g$ is defined in eq.(\ref{EOMMBCmassiveH}) with respect to $\bar{h}_{ij}^{(1)}$ instead of the induced metric $\cosh^2(\rho) \bar{h}_{ij}^{(1)}$. For the induced metric, the mass is defined by
 \begin{eqnarray}\label{EOMMassinducedmetric}
  \left(\Box+\frac{2}{\cosh^2(\rho)}-M_g^2\right)\bar{h}^{(1)}_{ij}(y)=0,
\end{eqnarray}
where $\Box=\bar{D}^i\bar{D}_i/\cosh^2(\rho)$ is the D'Alembert operator with respect to $\cosh^2(\rho) \bar{h}_{ij}^{(1)}$. Comparing eq.(\ref{EOMMBCmassiveH}) with eq.(\ref{EOMMassinducedmetric}), we read off
 \begin{eqnarray}\label{Massandmassgravity}
M_g^2=\frac{m_g^2}{\cosh^2(\rho)}.
\end{eqnarray}
For the large $\rho$, the mass spectrum of $M_g^2$ becomes almost continuous. While for the small $\rho$, the massive modes are frozen at low energy due to the infinite masses and only the massless mode is excited. As a result, the effective theory of wedge holography at low energy is Einstein gravity. 

\begin{figure}[t]
\centering
\includegraphics[width=7.5cm]{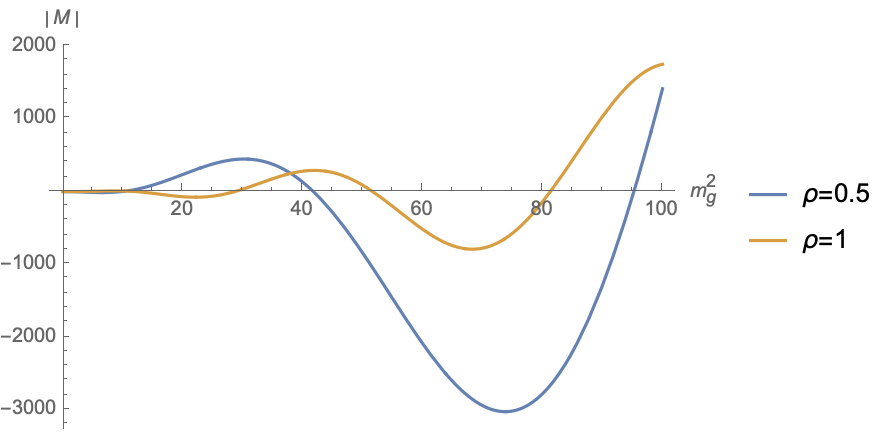}
\includegraphics[width=7.5cm]{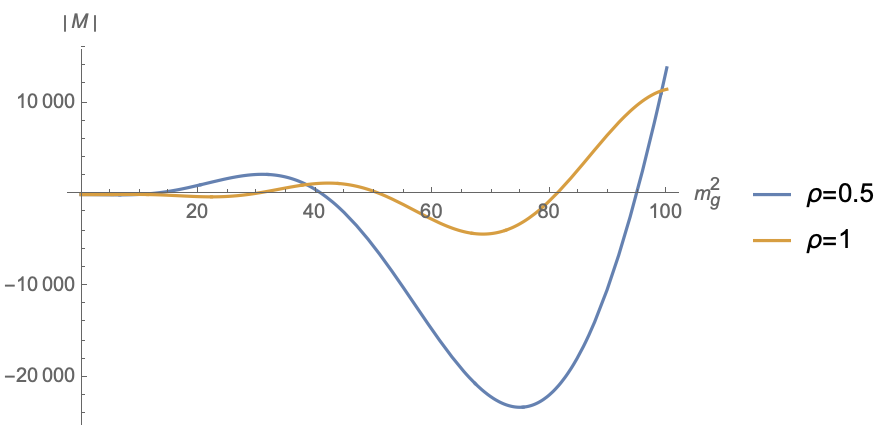}
\caption{Mass spectrum for gravitons with NBC. The left figure is for $d=3$, and the right figure is for $d=4$,  $|M|$ is given by eq.(\ref{constraintmassgravityNBC}) and the roots of $|M|=0$ denote the mass squares. The massless mode is always a solution to NBC. Besides, the larger the tension $T=(d-1)\tanh(\rho)$ is, the continuous the mass spectrum is. }
\label{gravitymass3d4dNBC}
\end{figure}

\begin{figure}[t]
\centering
\includegraphics[width=7.5cm]{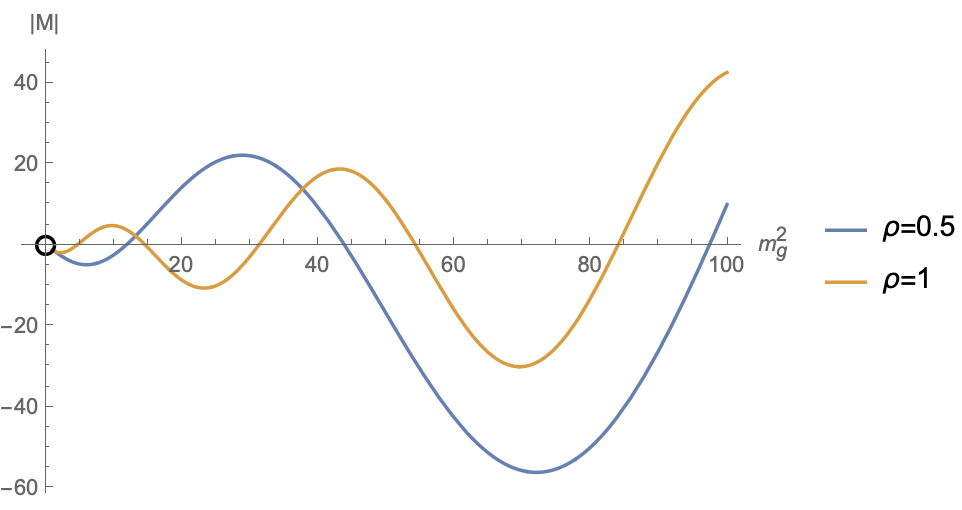}
\includegraphics[width=7.5cm]{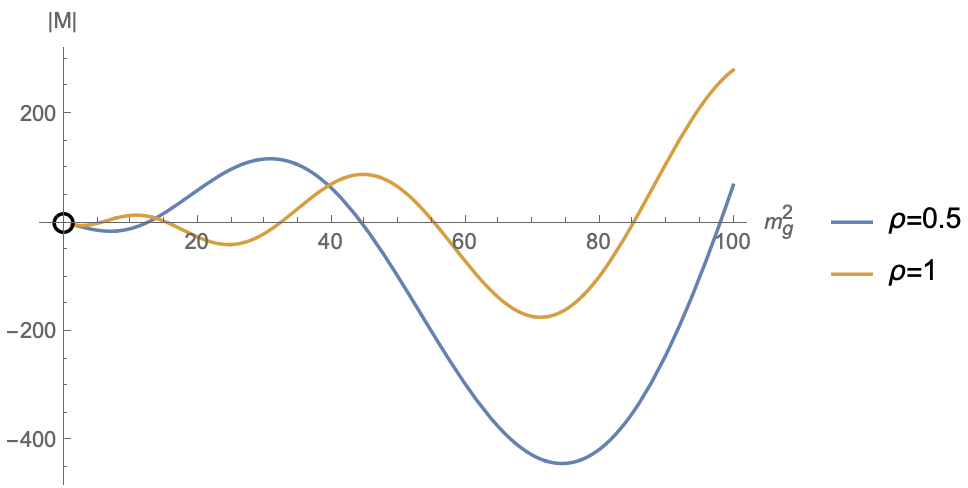}
\caption{Mass spectrum for gravitons with DBC. The left figure is for $d=3$, and the right figure is for $d=4$,  $|M|$ is given by eq.(\ref{constraintmassgravityDBCCBC}) and the roots of $|M|=0$ denote the mass squares. 
The small circle at origin means that the massless mode is removed from the mass spectrum for DBC. 
The larger the tension $T=(d-1)\tanh(\rho)$ is, the continuous the mass spectrum is.  }
\label{gravitymass3d4dDBC}
\end{figure}

{\bf 5.} Let us draw some figures to illustrate the spectrum.  See Fig.\ref{gravitymass3d4dNBC} and Fig.\ref{gravitymass3d4dDBC} for NBC and DBC respectively, where the intersections of the curves and $m_g^2$-axis denote the mass squares.  We notice that the massless mode is indeed a solution to NBC. For both NBC and DNC, the larger the tension $T=(d-1)\tanh(\rho)$ is, the continuous the mass spectrum is. 


\subsection{Perturbative action}
Let us go on to discuss the perturbative action of gravitons on the end-of-the-world branes.  Similarly, we expand the metric perturbations eq.(\ref{bulkmetricperturbations}) in powers of KK modes 
 \begin{eqnarray}\label{gravity-power}
\delta g_{r \mu}=0,  \ \ \delta g_{ij}=\cosh(r)^2 \sum_{m_g}H^{(m_g)}(r)\bar{h}^{(m_g)}_{ij}(y), 
\end{eqnarray}
where $\sum_{m_g}$ denotes the sum over the gravitational mass spectrum,  $ \bar{h}^{(m_g)}_{ij}(y)$ and $H^{(m_g)}(r)$ satisfy EOM eqs.(\ref{EOMMBCmassivehij},\ref{EOMMBCmassiveH}) and the orthogonal condition
 \begin{eqnarray}\label{orthogonal-gravity}
\int_{-\rho}^{\rho}\cosh(r)^{d-2}H^{(m_g)}(r) H^{(m'_g)}(r) dr=\delta^{m_g, m'_g}. 
\end{eqnarray}
The Einstein-Hilbert action expanded to second order of the metric perturbation is given by
 \begin{eqnarray}\label{gravityIbulk}
&&I_{EH}=\int_W d^{d+1}x\sqrt{|g|} \Big{[}-\frac{1}{4} \nabla_{\a} \bar{H}_{\mu\nu} \nabla^{\alpha} \bar{H}^{\mu\nu}+
\frac{1}{2} \nabla_{\alpha} \bar{H}_{\mu\nu} \nabla^{\nu} \bar{H}^{\mu \alpha}-\frac{d}{2}  \bar{H}_{\mu\nu}  \bar{H}^{\mu\nu} \Big{]} \nonumber\\
&&\ \ \ \ \ \ \ + \int_{\partial W}d^{d}y\sqrt{|h|} n^{\mu}\bar{H}^{\a\b} (\nabla_{\mu} \bar{H}_{\a\b}-\nabla_{\a} \bar{H}_{\b \mu} ),
 \end{eqnarray}
where we have set $16\pi G_N=1$, $\bar{H}_{\mu\nu} =\delta g_{\mu\nu}$ denote the metric perturbation and we have used the gauge eq.(\ref{gijgauge}) for simplicity.  The second order of perturbations of the Gibbons-Hawking term plus the brane action are
 \begin{eqnarray}\label{gravityIbdyQ}
I_{GH}=2\int_{\partial W} d^{d}y\sqrt{|h|} (K-T)=\int_{\partial W} d^{d}y\sqrt{|h|} \tanh(\rho) (-\frac{1}{2} \bar{H}_{\a\b}  \bar{H}^{\a\b}),
 \end{eqnarray}
 where we have used $K-T=\tanh(\rho)$, which applies to NBC and CBC at all orders, to DBC at the linear order.  For the metric ansatz eq.(\ref{perturbationmetric}) and the gauge eq.(\ref{gijgauge}), we have $\nabla_r \bar{H}_{ij}|_{\partial W}=0$ and $\nabla_i \bar{H}_{jr}=-\tanh(r) \bar{H}_{ij}$, which yield 
 \begin{eqnarray}\label{gravityformula}
n^{\mu}\bar{H}^{\a\b} \nabla_{\mu} \bar{H}_{\a\b}|_{\partial W}=0, \ \ n^{\mu}\bar{H}^{\a\b}\nabla_{\a} \bar{H}_{\b \mu}|_{\partial W} =-\tanh(\rho) \bar{H}_{\a\b} \bar{H}^{\a\b}.
 \end{eqnarray}
 From eqs.(\ref{gravityIbulk},\ref{gravityIbdyQ},\ref{gravityformula}), we obtain the total action
  \begin{eqnarray}\label{gravityItotal}
&&I=\int_W d^{d+1}x \sqrt{|g|} \Big{[}-\frac{1}{4} \nabla_{\a} \bar{H}_{\mu\nu} \nabla^{\alpha} \bar{H}^{\mu\nu}+
\frac{1}{2} \nabla_{\alpha} \bar{H}_{\mu\nu} \nabla^{\nu} \bar{H}^{\mu \alpha}-\frac{d}{2}  \bar{H}_{\mu\nu}  \bar{H}^{\mu\nu} \Big{]} \nonumber\\
&&\ \ \ \ \ +\frac{1}{2} \int_{\partial W} d^{d}y  \sqrt{|h|} \tanh(\rho) \bar{H}_{\a\b}  \bar{H}^{\a\b}.
 \end{eqnarray}
Substituting eq.(\ref{gravity-power}) into eq.(\ref{gravityItotal}) and following the approach of vectors, we finally derive
 \begin{eqnarray}\label{FPaction}
&&I=\sum_{m_g} \int_{Q}d^{d}y \sqrt{|\bar{h}^{(0)}|} \Big{[}-\frac{1}{4} \bar{D}_{k} \bar{h}^{(m_g)}_{ij} \bar{D}^{k} \bar{h}^{(m_g)}{}^{ij} +
\frac{1}{2} \bar{D}_{k} \bar{h}^{(m_g)}_{ij} \bar{D}^{i} \bar{h}^{(m_g)}{}^{jk}\nonumber\\
&&\ \ \ \ \ \ \ \ \ \ \ \ \ \ \ \ \ \ \ -\frac{d-1}{2} \bar{h}^{(m_g)}_{ij}  \bar{h}^{(m_g)}{}^{ij} -\frac{1}{4} m_g^2  \bar{h}^{(m_g)}_{ij}  \bar{h}^{(m_g)}{}^{ij} \Big{]},
\end{eqnarray}
which is an infinite sum of the Fierz-Pauli action in 
$\text{AdS}_d$ with the gauge eq.(\ref{hij1gauge}). To derive eq.(\ref{FPaction}), we find the following formulas are useful
  \begin{eqnarray}\label{gravitygoodformula1}
&&\nabla_r \bar{H}_{ij}=\cosh^2(r) \sum_{m_g} \bar{h}^{(m_g)}_{ij}(y) \frac{d}{dr}H^{(m_g)}(r), \\ \label{gravitygoodformula2}
&& \nabla_i \bar{H}_{jr}= -\cosh(r)\sinh(r)\sum_{m_g} \bar{h}^{(m_g)}_{ij}(y)H^{(m_g)}(r),\\
\label{gravitygoodformula3}
&& \nabla_i \bar{H}_{jk}= \cosh^2(r)\sum_{m_g} \bar{D}_i\bar{h}^{(m_g)}_{jk}(y)H^{(m_g)}(r).
 \end{eqnarray}
 Note that the induced metric on the brane is $\cosh^2(\rho) H(\pm \rho) \bar{h}^{(m_g)}_{ij}$ instead of $\bar{h}^{(m_g)}_{ij}$. Taking into account this fact and the normalization eq.(\ref{orthogonal-gravity}), from eq.(\ref{FPaction}) we can recover the effective Newton's constant of \cite{Miao:2020oey}. 
 
Note that action eq.(\ref{FPaction}) has the correct sign of kinetic term, thus it is ghost-free.  This is consistent with the discussions of sect. 2.2, which shows that the higher derivative effective action on the brane is ghost-free. 


\section{First law of entanglement entropy}
In this section, we prove that the first law of entanglement entropy is satisfied for wedge holography.  Since the holographic formula of entanglement entropy is still un-known for wedge holography with CBC/DBC \cite{Chu:2021mvq} 
\footnote{ Recall that, for CBC/DBC, the dynamical gravity on the brane is the extrinsic curvature instead of the induced metric. So far we know little about this kind of gravity. What we know is that, the holographic entanglement entropy for CBC/DBC must be different from that of NBC, since they have different spectrum and effective action on the brane.  Two natural questions are as follows. First, is the holographic entanglement entropy in the bulk still given by RT surface (extremal surface) with respect to the metric? Second, if one could impose CBC/DBC for gravity, can one impose CBC/DBC for the RT surface, that the endpoint of RT surface is fixed on the brane?  See \cite{Ghosh:2021axl} for some discussions on the RT surface with DBC. We leave a careful study of these problems to future works.}, we focus on NBC in this  section. Interestingly, we find that the massive fluctuations do not change the holographic entanglement entropy. For simplicity, we focus on the first order perturbation including the massless and massive modes of gravity in this section. 

To start, let us give a brief review of the relative entropy and the first law of entanglement entropy. 
 The relative entropy measures the fundamental distance between two states in the same Hilbert space
 \begin{eqnarray}\label{relative entropy}
S(\rho_1|\rho_0)=\text{tr}(\rho_1\ln \rho_1)-\text{tr}(\rho_1\ln \rho_0).
\end{eqnarray}
It is non-negative, i.e., $S(\rho_1|\rho_0)\ge 0 $, and vanishes if and only if the states are equal.  The relative entropy can be re-expressed as 
 \begin{eqnarray}\label{relative entropy1}
S(\rho_1|\rho_0)=\Delta \langle H \rangle-\Delta S,
\end{eqnarray}
with
 \begin{eqnarray}\label{relative entropy2}
\Delta \langle H \rangle=\text{tr}(\rho_1 H)-\text{tr}(\rho_0 H),\ \ \Delta S=S(\rho_1)-S(\rho_0),
\end{eqnarray}
where $H$ is the modular Hamiltonian defined by $\rho_0=e^{-H}/\text{tr}(e^{-H})$, and  $S(\rho)=-\text{tr}(\rho\ln \rho)$ is the entanglement entropy. In general, the positivity of the relative entropy requires
 \begin{eqnarray}\label{relative entropy3}
\Delta \langle H \rangle\ge \Delta S,
\end{eqnarray}
which is a quantum generalization of the Bekenstein bound.  
At the first order of perturbations, the above inequality is saturated \cite{Blanco:2013joa}
 \begin{eqnarray}\label{relative entropy4}
\delta \langle H \rangle = \delta S.
\end{eqnarray}
This is the so-called first law of entanglement entropy. Interestingly, by applying the first law of entanglement entropy eq.(\ref{relative entropy4}), one can derive the linearized Einstein Equations in the bulk \cite{Faulkner:2013ica}.

In the followings, we focus on the first order perturbations and verify that the first law of entanglement entropy is obeyed by wedge holography.

\subsection{The massive fluctuations}

Let us first discuss the massive fluctuations. We take the ansatz of the metric \eqref{perturbationmetric}
\begin{eqnarray} \label{metricEE}
ds^{2} =  dr^{2}+\textrm{cosh}^{2}(r)\Big( \frac{dz^{2}+\eta_{ab} dy^a dy^b}{z^2}+ \e \sum_{m_{g}}H^{(m_{g})}(r)\bar{h}_{ab}^{(m_{g})}(y) dy^a dy^b\Big),
\end{eqnarray}
where $y^{i}=(z,y^{a}),y^{a}=(t,x^{A})=(t,x^{1},...,x^{d-2})$, the index $a$ runs from 0 to $(d-2)$ and $\sum_{m_{g}}$ denotes the sum over the spectrum. 

Consider the entanglement entropy of $\text{CFT}_{d-1}$ on a $(d-2)$ dimensional round disk $\sum_{A=1}^{d-2}(x^{A})^{2}\le R_0^{2}$, where $R_0$ is the radius of the disk.  According to \cite{Ryu:2006bv}, it can be calculated by the area of the RT surface in the bulk
\begin{eqnarray}\label{RTfirstlaw}
	S  =  \frac{\text{Area}(\gamma)}{4G_{N}},
\end{eqnarray}
where $\gamma$ is the minimal surface anchored at the entangling surface of the disk, i.e., $\partial \gamma=\sum_{A=1}^{d-2}(x^{A})^{2}=R_0^{2}$. For the background metric eq.(\ref{metricEE}) with $\e=0$, the RT surface in the bulk is given by \cite{Akal:2020wfl}
\begin{eqnarray}
\text{RT surface}\  \gamma:\ 	z^{2}+r_0^2 =R_0^{2}, \ t=\text{constant}. \label{eq:RT}
\end{eqnarray}
where $r_0^2=\sum_{A=1}^{d-2}(x^{A})^{2}$. Since the RT surface is a minimal surface, the first order perturbation does not change its location. As a result, the RT surface for the perturbative metric eq.(\ref{metricEE}) is still given by eq.(\ref{eq:RT}).  From eq.(\ref{metricEE}) and eq.(\ref{eq:RT}), we read off the induced metric on the RT surface 
\begin{eqnarray} \label{metriconRT}
	ds_{\gamma}^2 = dr^{2}+\frac{\textrm{cosh}^{2}(r)}{z^2} \gamma_{AB} dx^A dx^B,
\end{eqnarray}
where
\begin{eqnarray}\label{metriconRT1}
	\gamma_{AB}=\delta_{AB}+ \partial_{A}z\partial_{B}z+ \e \ z^2\sum_{m_{g}}H^{(m_{g})}(r)\bar{h}_{AB}^{(m_{g})}(y) .\end{eqnarray}
By applying the RT formula eq.(\ref{RTfirstlaw}), we get the holographic entanglement entropy of a disk 
\begin{eqnarray}\label{RTfirstlaw1}
	S   =  \frac{1}{4G_{N}}\int_{-\rho}^{\rho}dr\int_{r_0\le R_0}d^{d-2}x{}\frac{\cosh^{d-2}(r)}{z^{d-2}}\sqrt{|\gamma_{AB}|},
\end{eqnarray}
which yields the first order variation of entanglement entropy
\begin{eqnarray}\label{eq:dS}
	\delta S = \sum_{m_{g}} \frac{\e}{8G_{N}}\int_{-\rho}^{\rho}dr \cosh^{d-2}(r) H^{(m_g)}(r)\int_{r_0\le R_0}d^{d-2}x{} f^{(m_g)}(x^A),
\end{eqnarray}
with
\begin{eqnarray}
	f^{(m_g)}(x^A)=\frac{1}{z^{d-4}}\left(\delta^{CD}\bar{h}_{CD}^{(m_{g})}\sqrt{1+\delta^{AB}\partial_{A}z\partial_{B}z}-\frac{\bar{h}^{(m_{g})CD}\partial_{C}z\partial_{D}z}{\sqrt{1+\delta^{AB}\partial_{A}z\partial_{B}z}}\right). 
\end{eqnarray}
Here $\bar{h}^{(m_{g})CD}$ are raised by $\delta^{CD}$ and $z$ obeys eq.(\ref{eq:RT}).  In the above derivations, we have used the formula  $|g_{AB}+\partial_{A}z\partial_{B}z|=|g_{AB}|(1+g^{AB}\partial_{A}z\partial_{B}z)$ \cite{Blanco:2013joa}.

Recall the orthogonal condition \eqref{orthogonal-gravity} and the fact that $H^{(0)}(r)$ is a constant for the massless mode. Immediately, we get
\begin{eqnarray}
	\int_{-\rho}^{\rho}dr\cosh^{d-2}(r)H^{(m_{g})}(r)  \sim \int_{-\rho}^{\rho}dr\cosh^{d-2}(r)H^{(m_{g})}(r) H^{(0)}(r)  =  0, \label{eq:orthogonal-factor}
\end{eqnarray}
for the massive modes $m_g>0$.  As a result, the first order variation of entanglement entropy eq.(\ref{eq:dS}) vanishes for the massive fluctuations. 

The first order variation of the expectation value of the modular
Hamiltonian $\delta\langle H\rangle$ for a spherical entangling
surface is given by \cite{Blanco:2013joa}
\begin{eqnarray} \label{eq:dH-formula}
	\delta\langle H\rangle  =  \frac{\pi}{R_0}\int_{r_0\le R_0}d^{d-2}x\ {}z^{2}{}\delta\langle T_{00}\rangle,
\end{eqnarray}
where $T_{00}$ is the stress tensor of $\text{CFT}_{d-1}$. Applying $\text{AdS}_d/\text{CFT}_{d-1}$ on the brane, we get the holographic stress tensor \cite{deHaro:2000vlm}
\begin{eqnarray} \label{eq:holographicTij}
	\langle T_{ab}\rangle  =  \frac{d-1}{16\pi G_{N}^{(d)}}h_{ab}^{(d-1)},
\end{eqnarray}
where $G_{N}^{(d)}$ is the effective Newton's constant, and $h_{ab}^{(d-1)}$ is defined in the Fefferman-Graham (FG) expansion on the brane
\begin{eqnarray}  \label{eq:holographicTij}
	ds_Q^2=\frac{dz^2+(\eta_{ab}+ z^{d-1}h_{ab}^{(d-1)} +... ) dy^a dy^b}{z^2}.
\end{eqnarray}
Note that, for the massive modes, the FG expansion behaves as
\begin{eqnarray} \label{massiveTij}
	ds_Q^2=\frac{dz^2+(z^{d-1-\Delta} h_{ab}^{(d-1-\Delta)}+ z^{\Delta }h_{ab}^{(\Delta)} +... ) dy^a dy^b}{z^2},
\end{eqnarray}
where $\Delta=\frac{d-1}{2}+\sqrt{\frac{(d-1)^{2}}{4}+m_{g}^{2}}$ is the conformal dimension, which can be derived from EOM eq.(\ref{EOMMBCmassivehij}).  Comparing eq.(\ref{eq:holographicTij}) with eq.(\ref{massiveTij}), we get $\langle T_{ab}\rangle \sim h_{ab}^{(d-1)}=0$  for the massive modes.  As a result, the first order variation of the expectation value of the modular Hamiltonian eq.(\ref{eq:dH-formula}) vanishes too.  Now we have verified the first law of entanglement entropy for the massive fluctuations
\begin{eqnarray} \label{firstlawmassive}
	\delta S=\delta \langle H \rangle =0, \  \ \text{for} \ m_g^2>0.
\end{eqnarray}

\subsection{The massless fluctuations}

Let us go on to discuss the massless fluctuations,  where $\bar{h}_{ab}^{(m_{g}=0)}(y^i)$ of
eq.(\ref{metricEE}) takes the form,
\begin{eqnarray} \label{eq:masslessmode}
	\bar{h}_{ab}^{(m_{g}=0)} (y^i) = z^{d-3}h_{ab}^{(d-1)}(y^a)=\frac{16\pi G_{N}^{(d)}}{d-1}z^{d-3}\langle T_{ab}(y^a)\rangle,
\end{eqnarray}
and the effective Newton's constant is given by \cite{Miao:2020oey}
\begin{eqnarray} \label{eq:GdN}
\frac{1}{G_{N}^{(d)}}=\frac{1}{G_{N}}\int_{-\rho}^{\rho}dr\cosh^{d-2}(r).
\end{eqnarray}
Note that we have set $H^{(m_{g}=0)}(r)=1$ for simplicity. Combining equations eq.(\ref{eq:RT},\ref{eq:dS},\ref{eq:masslessmode}), we get the first order perturbation of entanglement entropy
\begin{eqnarray}
	\delta S =  \frac{2\pi R_0S_{d-4}}{d-1}\int_{0}^{R_0}dr_0r_0^{d-3}\int_{0}^{\pi}d\theta \textrm{sin}^{d-4}\theta\left(\langle T_{\;\;\;A}^{A}(y)\rangle-\langle T_{AB}(y)\rangle\frac{x^{A}x^{B}}{R_0^{2}}\right),\label{eq:dSmassless}
\end{eqnarray}
where $S_{d-4}$ is the volume of $(d-4)$ dimensional unit sphere.  On the other hand, the first order perturbation of the expectation value of the modular
Hamiltonian is eq.(\ref{eq:dH-formula})
\begin{align} \label{eq:dHmassless}
	\delta\langle H\rangle & =  \frac{\pi}{R_0}\int_{r_0\le R_0}d^{d-2}x\ z^{2}\langle T_{00}(y)\rangle\nonumber \\
	& =  \frac{\pi S_{d-4}}{R_0}\int_{0}^{R_0}dr_0 r_0^{d-3}\int_{0}^{\pi}d\theta\textrm{sin}^{d-4}\theta(R_0^{2}-r_0^{2})\langle T_{00}(y)\rangle.
\end{align}

It is convenient to express $\delta S,\delta\langle H\rangle$ eqs.(\ref{eq:dSmassless},\ref{eq:dHmassless}) in the Fourier expansion
\begin{eqnarray}\label{eq:FT}
	T_{ab}(y) =  \int d^{d-1}pe^{-ip\cdot y}\hat{T}_{ab}(p). 
\end{eqnarray}
Without loss of generality, we drop the integral $ \int d^{d-1}pe^{ip^0 t}$ below. 
Following  \cite{Blanco:2013joa}, we set the spatial direction of momentum in direction $x^{1}$. The conservation and tracelessness of $\langle T_{ab}\rangle$ yield the following useful formulas \cite{Blanco:2013joa} 
\begin{align}\label{eq:TiiminusTij}
\langle\hat{T}_{\;\;\;A}^{A}\rangle=\langle\hat{T}_{00}\rangle,\;\; \; &\langle\hat{T}_{10}\rangle=-\frac{p^{0}}{p^{1}}\langle \hat{T}_{00}\rangle,\;\; \; \langle\hat{T}_{11}\rangle=\left(\frac{p^{0}}{p^{1}}\right)^{2}\langle\hat{T}_{00}\rangle,\nonumber\\
    \langle\hat{T}_{\;\;\;A}^{A}(p)\rangle-\langle\hat{T}_{AB}(p)\rangle\frac{x^{A}x^{B}}{R_0^{2}} & =  \langle\hat{T}_{00}\rangle \left(1-(\frac{p^{0}x^{1}}{p^{1}R_0})^{2}-\frac{(1-(\frac{p^{0}}{p^{1}})^{2})(r_0^2-(x^{1})^{2})}{(d-3)R_0^{2}}\right)\nonumber \\
 &= \langle\hat{T}_{00}\rangle\left(\frac{R_0^{2}-r_0^2\textrm{cos}^{2}(\theta)}{R_0^{2}}+\mathcal{O}(p^{2})\right), 
\end{align}
where $p^{0}=\sqrt{\text{(}p^{1})^{2}-p^{2}}$.

Substituting eqs.(\ref{eq:FT},\ref{eq:TiiminusTij}) into eqs.(\ref{eq:dSmassless},\ref{eq:dHmassless}), we finally derive
\begin{align}\label{eq:dSlast}
	\delta S & =  \frac{2\pi R_0S_{d-4}}{d-1}\langle\hat{T}_{00}\rangle \int_{0}^{R_0}dr_0 r_0^{d-3}\int_{0}^{\pi}d\theta \textrm{sin}^{d-4}\theta e^{-ip^{1}r_0\textrm{cos(\ensuremath{\theta})}}\left(\frac{R_0^{2}-r_0^2\textrm{cos}^{2}(\theta)}{R_0^{2}}\right)\nonumber \\
	& =  \frac{\pi^{\frac{3}{2}}R_0^{d-1}S_{d-4}\Gamma(\frac{d-3}{2})}{2\Gamma(\frac{d+2}{2})}{}_{0}F_{1}\left(\frac{d+2}{2};-\frac{1}{4}(p^{1}R_0)^{2}\right) \langle\hat{T}_{00}\rangle,
\end{align}
and
\begin{align}\label{eq:dHlast}
	\delta\langle H\rangle & =  \frac{\pi S_{d-4}}{R_0}\langle\hat{T}_{00}\rangle \int_{0}^{R_0}dr_0 r_0^{d-3}\int_{0}^{\pi} d\theta \textrm{sin}^{d-4}\theta e^{-ip^{1} r_0 \textrm{cos(\ensuremath{\theta})}}(R_0^{2}-r_0^{2})\nonumber \\
	& =  \frac{\pi^{\frac{3}{2}}R_0^{d-1}S_{d-4}\Gamma(\frac{d-3}{2})}{2\Gamma(\frac{d+2}{2})}{}_{0}F_{1}\left(\frac{d+2}{2};-\frac{1}{4}(p^{1}R_0)^{2}\right) \langle\hat{T}_{00}\rangle,
\end{align}
where $_{0}F_{1}\left(a;z\right)$ is the hypergeometric function. As expected, eqs.(\ref{eq:dSlast},\ref{eq:dHlast}) yield the correct first law of entanglement entropy $\delta S=\delta\langle H\rangle$. It is a strong support for wedge holography.

To end this section, let us make some comments. First, for the massive fluctuations, the first law of entanglement entropy eq.(\ref{firstlawmassive}) holds for more general background instead of only AdS.  That is because the EOM eq.(\ref{EOMMBCmassiveH}), the spectrum and the orthogonal condition eq.(\ref{orthogonal-gravity}) of $H(r)$ are independent of the choices of the background metric $h^{(0)}_{ij}$, as long as it obeys the Einstein equations on the brane.  As a result, $\delta S$ eq.(\ref{eq:dS}) always includes a vanishing pre-factor $\int_{-\rho}^{\rho}dr\cosh^{d-2}(r)H^{(m_{g})}(r)=0$ for the massive fluctuations. Similarly, we always have $\delta \langle H\rangle=0$ for the massive fluctuations, since the massive modes are irrelevant to the holographic stress tensor $\langle T_{ab}\rangle$ with the conformal dimension $(d-1)$. Thus, we can prove the first law of entanglement entropy $\delta S= \delta \langle H\rangle=0$ for the massive fluctuations around more general background. Second, it is straightforward to generalize the discussions of the massless modes to higher order perturbations.
Recall that the effective theory of the massless modes is Einstein gravity eq.(\ref{IWEinstein}) on the brane \cite{Miao:2020oey}.   According to \cite{Blanco:2013joa}, for Einstein gravity, we have $\delta S\le \delta \langle H\rangle$ at the second order perturbations. 
Third, inspired by \cite{Faulkner:2013ica}, it is interesting to turn the logic around, and to derive the Einstein equations in the $(d+1)$  dimensional wedge space from the first law of entanglement entropy of CFTs on the $(d-1)$ dimensional corner of the wedge.

\section{Conclusions and Discussions}

In this paper, we have investigated  the effective action, the spectrum and the first law of entanglement entropy for wedge holography. We work out the effective action on the brane. In the perturbative formulation, it is given by an infinite sum of Pauli-Fierz actions. In the non-perturbative formulation, the effective action is composed of a higher derivative gravity and a matter action. Usually, a higher derivative gravity suffers the problem of ghost. Due to the equivalence to Einstein gravity in the bulk, the effective action on the brane must be ghost-free. We find that the matter action plays an important role in eliminating the ghost. Besides, we provide evidences that the higher derivative gravity on the brane is equivalent to a ghost-free multi-gravity \cite{Hinterbichler:2012cn,deRham:2014zqa}. We show that the higher derivative gravity gives the correct Weyl anomaly of wedge holography, which is a support to both wedge holography and the non-perturbative effective action on the brane.  We also study the mass spectrum for various boundary conditions. We find that the spectrum is discrete and non-negative for all kinds of boundary conditions. In particular, there is a massless mode if one imposes NBC on both branes. On the other hand, the massless mode disappears if one imposes CBC/DBC on one or two of the branes. Finally, we verify that the first law of entanglement entropy is satisfied for wedge holography. Interestingly,  the massive fluctuations are irrelevant to the first order perturbations of entanglement entropy. This is also a strong support for wedge holography. 

In this paper, we mainly focus on NBC. It is interesting to discuss the 
effective action and the first law of entanglement entropy for CBC/DBC \cite{Miao:2018qkc,Chu:2021mvq}. It is expected that, similar to the case of NBC, the non-perturbative effective action for CBC/DBC is also equivalent to a higher derivative gravity or a multi-gravity on the brane. The only difference are the parameters of the theory, since, as we have shown in sect.3, the perturbative effective action is given by an infinite sum of Pauli-Fierz massive gravity with different mass spectrums for different BCs. It is also interesting to study the holographic entanglement entropy for gravity with CBC/DBC. Recall that the dynamical gravity on the brane is the extrinsic curvature instead of the induced metric. Is the holographic entanglement entropy still given by the area of a minimal surface with respect to the metric in the bulk? 
Which boundary condition should one choose for the RT surface ending on the brane with CBC/DBC? 
Finally, inspired by \cite{Faulkner:2013ica}, it is interesting to derive the Einstein equations in the $(d+1)$  dimensional wedge space from the first law of entanglement entropy on the $(d-1)$ dimensional corner of the wedge. We leave a 
careful study of these interesting problems to future works.

\section*{Acknowledgements}

We thank Chong-Sun Chu and Jie Ren  for valuable discussions. This work is supported by the National Natural Science Foundation of China (Grant No.11905297) and Guangdong Basic and Applied Basic Research Foundation (No.2020A1515010900).

\appendix

\section{Multi-gravity on the brane}

\begin{figure}[t]
\centering
\includegraphics[width=9cm]{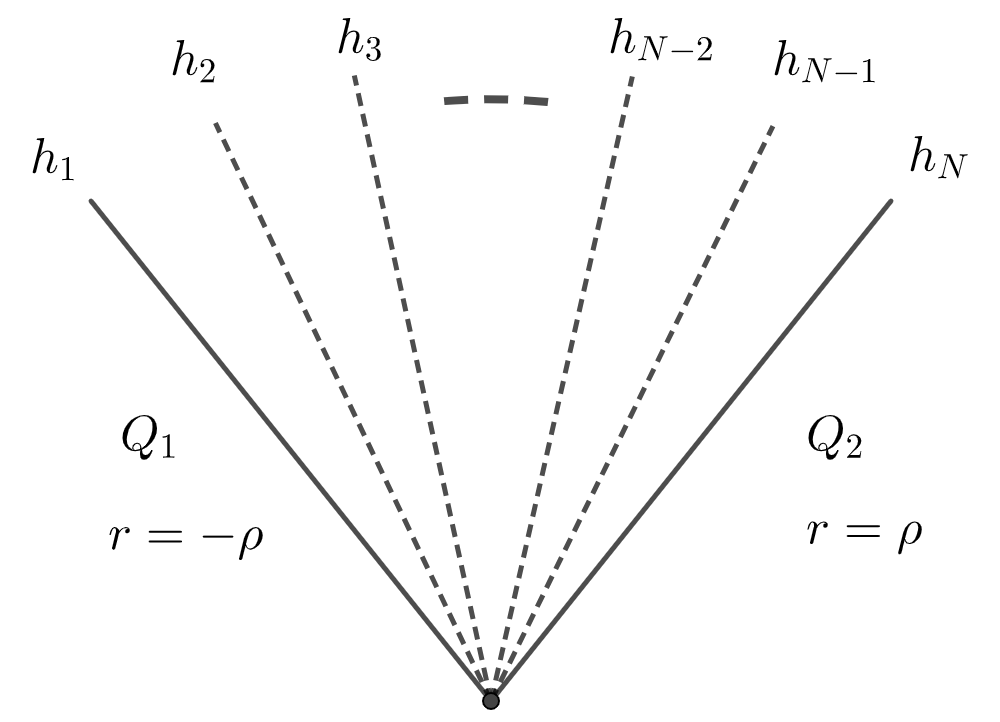}
\caption{Schematic diagram of deconstruction. The main idea is
to make discretization of Einstein gravity and replace the extra dimension $r$ by a series of sites $r_{b}$ ($1\le b \le N$). In this way, one can derive ghost-free multi-gravity from the Einstein gravity in the bulk. }
\label{multigravity}
\end{figure}

By applying the ``deconstruction'' method \cite{deRham:2014zqa}, one can obtain the so-called multi-gravity from Einstein gravity in the bulk.  The main idea is
to make discretization of Einstein gravity and replace the extra dimension $r$ by a series of sites $r_{b}$ ($1\le b \le N$). See Fig. \ref{multigravity} for an example.  
The Einstein gravity in the bulk can be written as
\begin{eqnarray} \label{app:Einsteingravity}
I=\int_{-\rho}^{\rho} dr \int dy^d\sqrt{|h|}  \Big(\mathcal{R}[h]+ K^2- K_{ij} K^{ij}-2\Lambda \Big),
\end{eqnarray}
where we have set $16\pi G_N=1$ for simplicity, $\mathcal{R}[h]$ is the intrinsic Ricci scalar on the constant $r$ surface, $K_{ij}=\frac{1}{2}\partial_r g_{ij}$ is the extrinsic curvature and $\Lambda$ is the cosmological constant. 
By discretization, the extrinsic curvature becomes 
\begin{eqnarray} \label{app:Kij}
&&K_{i}^a=e^{j a} K_{ij}\backsimeq m_N (e_{n+1}\ {}^a_i-e_{n}\ {}^a_i),\nonumber\\
&& K^i_j[h_n, h_{n+1}]=-m_N\left(\delta^i_j-\left(\sqrt{h_n^{-1}h_{n+1}}\right)^i_j\right),
\end{eqnarray}
where $e_{i a}$ is the vielbein and $m_N=N/(2\rho)\backsimeq \partial_r$ denotes the discrete derivative. 
The integration over the whole of the extra dimension can be replaced by summing over all the sites after discretization. In this way, we get the multi-gravity  eq.(\ref{multi-HDgravity})  \cite{Hinterbichler:2012cn,deRham:2014zqa}
\begin{eqnarray} \label{app:multigravity}
I&=&\sum_{n=1}^N \int dy^d\sqrt{|h_n|}  \left(\mathcal{R}[h_n]+  K^2[h_n, h_{n+1}] - K_{ij}[h_n, h_{n+1}] K^{ij}[h_n, h_{n+1}]-2\Lambda  \right)\nonumber\\
&=&  \sum_{n=1}^{N}\int dy^d  \sqrt{|h_n|} \left( \mathcal{R}_n +\frac{m_N^2}{2} \sum_{m=0}^{d}\alpha_m^{(n)} \mathcal{L}_m\left(K(h_n, h_{n+1})\right)\right),
\end{eqnarray}
where $\alpha_m^{(n)} $ are some coefficients, $\mathcal{L}_m[Q]=\epsilon \epsilon Q^m$ denote the interaction between neighboring metrics and $\epsilon$ is the 
Levi-Civita symbol.  For instance, we have $\mathcal{L}_2[Q]=\epsilon^{ijkl} \epsilon_{i_1j_1kl} Q^{i_1}_iQ^{j_1}_j$ in four dimensions. 
By construction, the multi-gravity eq.(\ref{app:multigravity}) is ghost-free \cite{Hinterbichler:2012cn,deRham:2014zqa}. 

Taking the limit $N\to \infty$, the multi-gravity eq.(\ref{app:multigravity})  reduces to Einstein gravity eq.(\ref{app:Einsteingravity}) in the bulk. 
  Since the multi-gravity eq.(\ref{app:multigravity}) with $N\to \infty$ and the effective higher derivative gravity eq.(\ref{IW}) on the two branes are both obtained from the bulk Einstein gravity, they must be equivalent. Following the approach of \cite{Hassan:2013pca}, by eliminating the metrics $h_2, h_3,..., h_{N-1}$ of the multi-gravity with $N\to \infty$, we should recover the higher derivative gravity eq.(\ref{IW}) on the branes. We leave a careful study of this problem to future work \cite{HuandMiao}. In this appendix, we consider finite $N$ for simplicity. We recover the correct form of $O(\mathcal{R}^2)$ terms in the higher derivative gravity eq.(\ref{IW}). This is a strong evidence for the equivalence between the higher derivative gravity on the brane and the ghost-free multi-gravity.  For simplicity, we focus on the symmetric deconstruction of the bulk gravity, 
where the action is invariant under the transformations $ (b)\leftrightarrow(N+1-b) $,
 so that we only need to consider half of the
wedge space 
$-\rho\le r\le0$ below. 

\subsection{Multi-gravity with four metrics}\label{A1}

Let us first discuss the multi-gravity with four metrics $h_{b}$ $(1\le b\le N=4)$.
The action includes three parts
\begin{equation}\label{I4}
	I_{4}(h_{b})=\int d^{d}y\left(\mathcal{L}_{12}+\mathcal{L}_{43}+\mathcal{L}_{int\;23}\right),
\end{equation}
where $\mathcal{L}_{12}$ and $\mathcal{L}_{43}$ are the bi-metric Lagrangian density for $(h_1, h_2)$ and $(h_4, h_3)$ respectively,  $\mathcal{L}_{int\;23}$ denotes the interaction of $(h_2, h_3)$
\begin{eqnarray}
	\label{L12}
	&&	\mathcal{L}_{12}  =m_{(1)}^{d-2}\left(\sqrt{|h_{1}|}\mathcal{R}_{1}+\gamma_{(2)}^{d-2}\sqrt{|h_{2}|}\mathcal{R}_{2}-2m^{2}\sqrt{|h_{1}|}\sum_{n=0}^{d}\beta_{n}^{(1)}e_{n}(S_{(1)})\right),\\
	\label{L43}
	&&	\mathcal{L}_{43}  =m_{(4)}^{d-2}\left(\sqrt{|h_{4}|}\mathcal{R}_{4}+\gamma_{(3)}^{d-2}\sqrt{|h_{3}|}\mathcal{R}_{3}-2m^{2}\sqrt{|h_{4}|}\sum_{n=0}^{d}\beta_{n}^{(4)}e_{n}(\bar{S}_{(4)})\right),\\
	&&	\mathcal{L}_{int\;23}  =-m^{2}\sum_{n=0}^{d}\left(m_{(1)}^{d-2}\beta_{n}^{(2)}\sqrt{|h_{2}|}e_{n}(S_{(2)})+m_{(4)}^{d-2}\beta_{n}^{(3)}\sqrt{|h_{3}|}e_{n}(\bar{S}_{(3)})\right).
\end{eqnarray}
Let us recall some notations. $\gamma_{(2)}=m_{(2)}/m_{(1)}$ and $\gamma_{(3)}=m_{(3)}/m_{(4)}$ are
the ratio of the Planck masses, $\beta_{n}^{(b)}$ are dimensionless
free parameters, $\mathcal{R}_{b}$ is the Ricci scalar of $h_{b}$,
$S_{(b)}=\sqrt{h_{b}^{-1}h_{b+1}}$ and $\bar{S}_{(b)}=\sqrt{h_{b}^{-1}h_{b-1}}$.
Note that $\mathcal{L}_{12}$ and $\mathcal{L}_{43}$ are symmetric
under the interchanges 
$\left\{ h_{b} \leftrightarrow  h_{N+1-b},(b)\leftrightarrow(N+1-b)\right\} $
because of the symmetric deconstruction of the bulk gravity. The equations
of motion (EOM) of $h_{1}$ and $h_{2}$ are given by
 \cite{Hassan:2013pca,Hassan:2012rq}
\begin{eqnarray}
	\label{h1EOMS1}
	&& \mathcal{P}_{1\;j}^{\;i}  -\mathcal{P}_{1}\delta_{j}^{i}+m^{2}\sum_{n=0}^{d-1}(-1)^{n}\beta_{n}^{(1)}\mathcal{\mathbb{Y}}_{(n)\nu}^{\mu}(S_{(1)})=0,\\
	\label{h2EOMS1S2}
	&& \mathcal{P}_{2\;j}^{\;i}  -\mathcal{P}_{2}\delta_{j}^{i}+\frac{m^{2}}{\gamma_{(2)}^{d-2}}\sum_{n=0}^{d-1}(-1)^{n}\left[\beta_{n}^{(2)}\mathcal{\mathbb{Y}}_{(n)\nu}^{\mu}(S_{(2)})+\beta_{d-n}^{(1)}\mathcal{\mathbb{Y}}_{(n)\nu}^{\mu}(S_{(1)}^{-1})\right]=0,
\end{eqnarray}
where $\mathcal{P}_{b\;ij}=\mathcal{R}_{b\;ij}-\frac{\mathcal{R}_{b}}{2(d-1)}h_{b\;ij}$,
$\mathcal{\mathbb{Y}}_{(n)\nu}^{\mu}(S_{(b)})=\sum_{k=0}^{n}(-1)^{k}e_{k}(S_{(b)})(S_{(b)}^{n-k})_{\;j}^{i}$
.  

We aim to express $h_2$ in functions of $h_1$. To do so, we first solve $S_{(1)\;j}^{\;i}=(\sqrt{h_1^{-1}h_2})^i_j$ perturbatively. The general ansatz of $S_{(1)\;j}^{\;i}$ is given by
\begin{align}\label{S1Ansatz}
	S_{(1)\;j}^{\;i}=&a^{(1)}\delta_{j}^{i} +  \frac{1}{m^{2}}\left(b_{1}^{(1)}\mathcal{P}_{1\;j}^{\;i}+b_{2}^{(1)}\mathcal{P}_{1}\delta_{j}^{i}\right)\nonumber \\
	& +  \frac{1}{m^{4}}\left(c_{1}^{(1)}\mathcal{P}_{1\;k}^{\;i}\mathcal{P}_{1\;j}^{\;k}+c_{2}^{(1)}\mathcal{P}_{1}\mathcal{P}_{1\;j}^{\;i}+c_{3}^{(1)}\mathcal{P}_{1\;kl}\mathcal{P}_{1}^{\;kl}\delta_{j}^{i}+c_{4}^{(1)}\mathcal{P}_{1}^{\;2}\delta_{j}^{i}\right)+\mathcal{O}(m^{-6}).
\end{align}
where $a^{(1)},b^{(1)}$ and $c^{(1)}$ are the coefficients to be determined below. Substituting eq.(\ref{S1Ansatz})
into eq.(\ref{h1EOMS1}), we get
\begin{align}\label{h1EOMM1}
	\mathcal{P}_{1\;j}^{\;i}-\mathcal{P}_{1}\delta_{j}^{i}+m^{2}\sum_{n=0}^{d-1}\beta_{n}^{(1)}a^{(1)n}\sum_{k=0}^{n}\left(\sum_{r=0}^{n-k}\sum_{m=0}^{k}(-1)^{n+k}C_{n-k}^{r}C_{d-m}^{k-m}(M_{(1)}^{r})_{\;j}^{i}e_{m}(M_{(1)})\right)=0,
\end{align}
where we have replaced $S_{(1)}$ by $M_{(1)}$ defined below
\begin{eqnarray}
	&&M_{(b)\;j}^{\;i}=\frac{1}{a^{(b)}}S_{(b)\;j}^{\;i}-\delta_{j}^{i},\label{app:MS}  \\
	&&(S_{(b)}^{n-k})_{\;j}^{i}=a^{(b)n-k}\sum_{r=0}^{n-k}(M_{(b)}^{r})_{\;j}^{i}, \ \ \  e_{k}(S_{(b)})=a^{(b)k}\sum_{m=0}^{k}C_{d-m}^{k-m}e_{m}(M_{(b)}).
\end{eqnarray}
From  eq.(\ref{S1Ansatz}) and  eq.(\ref{app:MS}), we have
\begin{align}\label{M1Ansatz}
	M_{(1)\;j}^{\;i}&= \frac{1}{m^{2}}\frac{1}{a^{(1)}}\left(b_{1}^{(1)}\mathcal{P}_{1\;j}^{\;i}+b_{2}^{(1)}\mathcal{P}_{1}\delta_{j}^{i}\right)\nonumber \\
	& +  \frac{1}{m^{4}}\frac{1}{a^{(1)}}\left(c_{1}^{(1)}\mathcal{P}_{1\;k}^{\;i}\mathcal{P}_{1\;j}^{\;k}+c_{2}^{(1)}\mathcal{P}_{1}\mathcal{P}_{1\;j}^{\;i}+c_{3}^{(1)}\mathcal{P}_{1\;kl}\mathcal{P}_{1}^{\;kl}\delta_{j}^{i}+c_{4}^{(1)}\mathcal{P}_{1}^{\;2}\delta_{j}^{i}\right)+\mathcal{O}(m^{-6}).
\end{align}
It's convenient to use the following notations and formulas
  \begin{align}
\label{sk} 	
  	&s_{k}^{(b)}=\sum_{n=k}^{d-1}C_{d-k-1}^{n-k}\beta_{n}^{(b)}a^{(b)n},\\
\label{BinomialEquation} 	
	&\sum_{k=0}^{n}(-1)^{k+n}C_{n-k}^r C_{d-m}^{k-m}=(-1)^rC_{d-1-(r+m)}^{n-(r+m)}.
\end{align}
 Combining eqs.(\ref{h1EOMM1},\ref{sk},\ref{BinomialEquation}),
  we obtain 
  \begin{align}\label{h1EOMExpansion}
	\mathcal{P}_{1\;j}^{\;i}-\mathcal{P}_{1}\delta_{j}^{i} & +  s_{0}^{(1)}m^{2}\delta_{j}^{i}+s_{1}^{(1)}m^{2}\left[-M_{(1)\;j}^{\;i}+\delta_{j}^{i}e_{1}(M_{(1)})\right]\nonumber \\
	& +  s_{2}^{(1)}m^{2}\left[(M_{(1)}^{2})_{\;j}^{i}-M_{(1)\;j}^{\;i}e_{1}(M_{(1)})+\delta_{j}^{i}e_{2}(M_{(1)})\right]+\mathcal{O}(M_{(1)}^{3})=0.
\end{align}
Substituting eq.(\ref{M1Ansatz}) into the above equation and solving it order by order in $O(1/m^2)$, we get
\begin{eqnarray}\label{S1Coefficients}
	\mathcal{O}(m^{2}), && s_{0}^{(1)}=0,\nonumber \\
	\mathcal{O}(m^{0}), && b_{1}^{(1)}=\frac{a^{(1)}}{s_{1}^{(1)}},\;\;\;b_{2}^{(1)}=0,\nonumber \\
	\mathcal{O}(m^{-2}), && c_{1}^{(1)}=-c_{2}^{(1)}=\frac{a^{(1)}s_{2}^{(1)}}{s_{1}^{(1)3}},\;\;\;c_{3}^{(1)}=-c_{4}^{(1)}=-\frac{a^{(1)}s_{2}^{(1)}}{2(d-1)s_{1}^{(1)3}}.
\end{eqnarray}
Substituting eq.(\ref{S1Coefficients}) into eq.(\ref{S1Ansatz}),
we finally obtain $S_{(1)}$
\begin{eqnarray}\label{S1Solution}
	S_{(1)\;j}^{\;i} =  a^{(1)}\delta_{j}^{i}+\frac{a^{(1)}}{s_{1}^{(1)}m^{2}}\mathcal{P}_{1\;j}^{\;i}+\frac{a^{(1)}s_{2}^{(1)}}{s_{1}^{(1)3}m^{4}}\left(\mathcal{P}_{1\;k}^{i}\mathcal{P}_{1\;j}^{k}-\mathcal{P}_{1}\mathcal{P}_{1\;j}^{i}+\frac{e_{2}(\mathcal{P}_{1})}{d-1}\delta_{j}^{i}\right)+\mathcal{O}(m^{-6}).
\end{eqnarray}
and express $h_2$ in terms of $h_1$
\begin{align}\label{h2Solution}
	h_{2\;ij} & =  h_{1\;ik}(S_{(1)}^{\;2})_{\;j}^{k}\nonumber \\
	& =  a^{(1)2}\left[h_{1\;ij}+\frac{2\mathcal{P}_{1\;ij}}{s_{1}^{(1)}m^{2}}+\frac{2s_{2}^{(1)}}{s_{1}^{(1)3}m^{4}}\left((1+\frac{s_{1}^{(1)}}{2s_{2}^{(1)}})\mathcal{P}_{1\;ik}\mathcal{P}_{1\;j}^{k}-\mathcal{P}_{1}\mathcal{P}_{1\;ij}+\frac{e_{2}(\mathcal{P}_{1})}{d-1}h_{1\;ij}\right)\right]\nonumber \\
	&\ \ +\mathcal{O}(m^{-6}).
\end{align}
Following the same approach, we can express $h_3$ in terms of $h_4$.  From $h_2(h_1)$ eq.(\ref{h2Solution}) and $h_3(h_4)$, the interaction potential $\mathcal{L}_{int\;23}(h_{2},h_{3})$
can be converted to $\mathcal{L}_{int\;14}(h_{1},h_{4})$. We find the following relations are useful in the calculations
\begin{align}\label{LintSToM}
	\sum_{n=0}^{d}\beta_{n}^{(b)}e_{n}(S_{(b)})=\sum_{n=0}^{d}\alpha_{n}^{(b)}e_{n}(M_{(b)}),\ \ \ & \alpha_{n}^{(b)}=\sum_{k=n}^{d}C_{k-n}^{d-n}a^{(b)k}\beta_{k}^{(b)}.
\end{align}

 Finally, combining eqs.(\ref{I4},\ref{L12},\ref{L43},\ref{S1Solution},\ref{h2Solution},\ref{LintSToM}),
we obtain the higher derivative gravity with respect to two metrics $h_{1}$ and $h_{4}$
\begin{align}\label{I4R2}
	I_{4}^{HD}(h_{1},h_{4})
	& =  m_{(1)}^{d-2}\int_{Q_{1}}d^{d}y \sqrt{|h_1|}\left[\hat{\Lambda}^{(1)}+\hat{c}_{\mathcal{R}}^{(1)}\mathcal{R}_{1}+\frac{\hat{c}_{\mathcal{R}\mathcal{R}}^{(1)}}{m^{2}}\left(\mathcal{R}_{1}^{\;ij}\mathcal{R}_{1\;ij}-\frac{d}{4(d-1)}\mathcal{R}_{1}^{\;2}\right)\right]\nonumber \\
	& +  m_{(4)}^{d-2}\int_{Q_{2}}d^{d}y \sqrt{|h_4|}\left[\hat{\Lambda}^{(4)}+\hat{c}_{\mathcal{R}}^{(4)}\mathcal{R}_{4}+\frac{\hat{c}_{\mathcal{R}\mathcal{R}}^{(4)}}{m^{2}}\left(\mathcal{R}_{4}^{\;ij}\mathcal{R}_{4\;ij}-\frac{d}{4(d-1)}\mathcal{R}_{4}^{\;2}\right)\right]\nonumber\\
	&+\mathcal{O}(\mathcal{R}_{1}^{\;3})+\mathcal{O}(\mathcal{R}_{4}^{\;3})+\mathcal{L}_{int\;14}(h_{1},h_{4}),
\end{align}
where the parameters are given by 
\begin{align}\label{I4R2Coefficients}
	\hat{\Lambda}^{(1)}&=-2m^{2}\alpha_{0}^{(1)},\ \ \ \hat{c}_{\mathcal{R}}^{(1)}=1+(\gamma_{(2)}a^{(1)})^{d-2}-\frac{\alpha_{1}^{(1)}(d-2)}{s_{1}^{(1)}(d-1)},\nonumber \\
	\hat{c}_{\mathcal{R}\mathcal{R}}^{(1)}&=-\frac{1}{s_{1}^{(1)2}}\left(2s_{1}^{(1)}(\gamma_{(2)}a^{(1)})^{d-2}-\alpha_{2}^{(1)}+\frac{\alpha_{1}^{(1)}s_{2}^{(1)}(d-2)}{s_{1}^{(1)}(d-1)}\right),
\end{align}
and $\hat{\Lambda}^{(4)},\hat{c}_{\mathcal{R}}^{(4)}, \hat{c}_{\mathcal{RR}}^{(4)}$
can be obtained from eq.(\ref{I4R2Coefficients}) by replacing indexes $(b)$ by $(5-b)$. 

Remarkably, the higher derivative gravity eq.(\ref{I4R2}) reduced from the multi-gravity takes similar forms as the effective action eq.(\ref{IW}) on the brane, where  the interaction $\mathcal{L}_{int\;14}(h_{1},h_{4})$ can be naturally explained as the CFT action $I_{\text{CFT}}$.

\subsection{Multi-gravity with six metrics}

Let us now discuss the multi-gravity with six metrics $h_{b}$ $(1\le b \le N=6)$
\begin{equation}\label{I6}
	I_{6}(h_{b})=\int d^{d}y\left(\mathcal{L}_{123}+\mathcal{L}_{654}+\mathcal{L}_{int\;34}\right),
\end{equation}
where $\mathcal{L}_{123}$ and $\mathcal{L}_{654}$ are the Lagrangian density of multi-gravity  with three metrics for $(h_1,h_2,h_3)$ and $(h_6,h_5,h_4)$ respectively, $\mathcal{L}_{int \; 34}$ denotes the interaction of $(h_3,h_4)$
\begin{align}\label{L6}
&	\mathcal{L}_{123}  =\mathcal{L}_{12}+\mathcal{L}_{3}=\mathcal{L}_{12}+m_{(1)}^{d-2}\left(\gamma_{(3)}^{d-2}\sqrt{|h_{3}|}\mathcal{R}_{3}-2m^{2}\sqrt{|h_{2}|}\sum_{n=0}^{d}\beta_{n}^{(2)}e_{n}(S_{(2)})\right), \\
&	\mathcal{L}_{654}  =\mathcal{L}_{65}+\mathcal{L}_{4}=\mathcal{L}_{65}+m_{(6)}^{d-2}\left(\gamma_{(4)}^{d-2}\sqrt{|h_{4}|}\mathcal{R}_{4}-2m^{2}\sqrt{|h_{5}|}\sum_{n=0}^{d}\beta_{n}^{(5)}e_{n}(\bar{S}_{(5)})\right), \\
&	\mathcal{L}_{int\;34}  =-m^{2}\sum_{n=0}^{d}\left(m_{(1)}^{d-2}\beta_{n}^{(3)}\sqrt{|h_{3}|}e_{n}(S_{(3)})+m_{(6)}^{d-2}\beta_{n}^{(4)}\sqrt{|h_{4}|}e_{n}(\bar{S}_{(4)})\right),
\end{align}
where $\mathcal{L}_{12}$ is given by eq.(\ref{L12}), $\mathcal{L}_{65}$
can be obtained from eq.(\ref{L43}) by replacing indexes (4,3) with (6,5).
Since we have already 
rewritten $\mathcal{L}_{12}$ in functions of $h_1$ in
 eq.(\ref{A1}), we focus on $\mathcal{L}_{3}$ below. 
 
 In order to express $h_3$ in functions of $h_1$, we first solve $S_{(2)\;j}^{\;i}=(\sqrt{h_2^{-1}h_3})^i_j$ perturbatively. The general
ansatz of $S_{(2)\;j}^{\;i}$ is given  by 
 \begin{align}\label{S2Ansatz}
 	S_{(2)\;j}^{\;i}=&\ a^{(2)}\delta_{j}^{i} +  \frac{1}{m^{2}}\left(b_{1}^{(2)}\mathcal{P}_{1\;j}^{\;i}+b_{2}^{(2)}\mathcal{P}_{2}\delta_{j}^{i}\right)\nonumber \\
 	+&\frac{1}{m^{4}}\left(c_{1}^{(2)}\mathcal{P}_{2\;k}^{\;i}\mathcal{P}_{2\;j}^{\;k}+c_{2}^{(2)}\mathcal{P}_{2}\mathcal{P}_{2\;j}^{\;i}+c_{3}^{(2)}\mathcal{P}_{2\;kl}\mathcal{P}_{2}^{\;kl}\delta_{j}^{i}+c_{4}^{(2)}\mathcal{P}_{2}^{\;2}\delta_{j}^{i}\right)+\mathcal{O}(m^{-6}),
 \end{align}
 where $a^{(2)}, b^{(2)}$ and $c^{(2)}$ are the coefficients to be determined below. Substituting  \eqref{S2Ansatz} into eq.(\ref{h2EOMS1S2}), we get
 \begin{align}\label{h2EOMM2M1}
 	\mathcal{P}_{2\;j}^{\;i}- & \mathcal{P}_{2}\delta_{j}^{i}+\text{\ensuremath{\frac{m^{2}}{\gamma_{2}^{d-2}}}}\sum_{n=0}^{d-1}\beta_{n}^{(2)}a^{(2)n}\sum_{k=0}^{n}\left(\sum_{r=0}^{n-k}\sum_{m=0}^{k}(-1)^{n+k}C_{n-k}^{r}C_{d-m}^{k-m}(M_{(2)}^{r})_{\;j}^{i}e_{m}(M_{(2)})\right)\nonumber \\
 	+ & \ensuremath{\frac{m^{2}}{\gamma_{2}^{d-2}}}\sum_{n=0}^{d-1}\beta_{d-n}^{(1)}a^{(1)-n}\sum_{k=0}^{n}\left(\sum_{r=0}^{n-k}\sum_{m=0}^{k}(-1)^{n+k}C_{n-k}^{r}C_{d-m}^{k-m}(\hat{M}_{(1)}^{r})_{\;j}^{i}e_{m}(\hat{M}_{(1)})\right)=0,
 \end{align}
 where $M_{(2)}$ is given by eqs.(\ref{app:MS},\ref{S2Ansatz}) and we have replaced $S_{(1)}^{-1}$ by $\hat{M}_{(1)}$ defined below
 \begin{align}
 \label{eq:MM}
 	&\hat{M}_{(b)\;j}^{\;i}=a^{(b)}(S_{(b)}^{-1})_{\;j}^{i}-\delta_{j}^{i},\\
 	&\left((S_{(b)}^{-1})^{n-k}\right)_{\;j}^{i}=a^{(b)k-n}\sum_{r=0}^{n-k}(\hat{M}_{(b)}^{r})_{\;j}^{i},\ \ \ e_{k}(S_{(b)}^{-1})=a^{(b)-k}\sum_{m=0}^{k}C_{d-m}^{k-m}e_{m}(\hat{M}_{(b)}).
 \end{align}
From eq.(\ref{S1Solution}) and eq.(\ref{eq:MM}), we have
\begin{equation}\label{M1hat}
	\hat{M}_{(1)\;j}^{\;i}= -\frac{\mathcal{P}_{1\;j}^{\;i}}{s_{1}^{(1)}m^{2}}+\frac{1}{s_{1}^{(1)3}m^{4}}\left(s_{2}^{(1)}\mathcal{P}_{1}\mathcal{P}_{1\;j}^{i}+(s_{1}^{(1)}-s_{2}^{(1)})\mathcal{P}_{1\;k}^{i}\mathcal{P}_{1\;j}^{k}-s_{2}^{(1)}\frac{e_{2}(\mathcal{P}_{1})}{d-1}\delta_{j}^{i}\right)+\mathcal{O}(m^{-6}).\\
\end{equation}
Combining eqs.(\ref{sk},\ref{BinomialEquation},\ref{h2EOMM2M1}), we obtain
\begin{align}\label{h2EOMExpansion}
\mathcal{P}_{2\;j}^{\;i}
-&\mathcal{P}_{2}\delta_{j}^{i}+\frac{m^{2}}{\gamma_{2}^{d-2}}\left[s_{0}^{(2)}\delta_{j}^{i}+s_{1}^{(2)}\left(-M_{(2)\;j}^{\;i}+\delta_{j}^{i}e_{1}(M_{(2)})\right)\right]\nonumber\\
+&\frac{m^{2}}{\gamma_{2}^{d-2}}\left[s_{2}^{(2)}\left((M_{(2)}^{2})_{\;j}^{i}-M_{(2)\;j}^{\;i}e_{1}(M_{(2)})+\delta_{j}^{i}e_{2}(M_{(2)})\right)\right]\nonumber\\
+&\frac{m^{2}}{\gamma_{2}^{d-2}}\left[q_{2}^{(1)}\left((\hat{M}_{(1)}^{2})_{\;j}^{i}-\hat{M}_{(1)\;j}^{\;i}e_{1}(\hat{M}_{(1)})+\delta_{j}^{i}e_{2}(\hat{M}_{(1)})\right)\right]\nonumber\\
+&\frac{m^{2}}{\gamma_{2}^{d-2}}\left[q_{0}^{(1)}\delta_{j}^{i}+q_{1}^{(1)}\left(-\hat{M}_{(1)\;j}^{\;i}+\delta_{j}^{i}e_{1}(\hat{M}_{(1)})\right)\right]+\mathcal{O}(\hat{M}_{(1)}^{3},M_{(2)}^{3})
=0,
\end{align}
where $q_{k}^{(b)}=\sum_{n=k}^{d-1}C_{d-k-1}^{n-k}\beta_{d-n}^{(b)}a^{(b)-n}$. 
Note that $\mathcal{P}_{2}$ in eq.(\ref{h2EOMExpansion}) can be
converted to $\mathcal{P}_{1}$ by eq.(\ref{h2Solution})
\begin{align}\label{P2ToP1}
	\mathcal{P}_{2\;j}^{\;i}-\mathcal{P}_{2}\delta_{j}^{i}=&\frac{1}{a^{(1)2}}\left[\mathcal{P}_{1\;j}^{\;i}-\mathcal{P}_{1}\delta_{j}^{i}+\frac{1}{s_{1}^{(1)}m^{2}}\left((\mathcal{P}_{1}^{\;kl}\mathcal{P}_{1\;kl}+\frac{\mathcal{P}_{1}^{\;2}}{d-2})\delta_{j}^{i}-2(\mathcal{P}_{1\;k}^{i}\mathcal{P}_{1\;j}^{k}+\frac{\mathcal{P}_{1}\mathcal{P}_{1\;j}^{i}}{d-2})\right)\right]\nonumber \\
   &+\mathcal{O}(m^{-4}),
\end{align}
where we have ignored the total derivative terms of $\mathcal{P}_1$, which have no contribution to 
the gravitational action at order $O(\mathcal{R}^2)$ \footnote{At order $O(\mathcal{R}^2)$, the only possible terms including derivatives are $\Box \mathcal{R}$ and $\nabla_i \nabla_j \mathcal{R}^{ij}$, which are total derivative terms and can be dropped in the action.}.  
 Substituting eqs.(\ref{M1hat},\ref{P2ToP1}) and $M_{(2)}$ into eq.(\ref{h2EOMExpansion}) and solving it order by order in $\mathcal{O}(1/m^2)$, we
obtain the coefficients of $S_{(2)\;j}^{\;i}$
\begin{align}\label{S2Coefficients}
	\mathcal{O}&(m^{2}),\ \   s_{0}^{(2)}=-q_{0}^{(1)},\nonumber \\
	\mathcal{O}&(m^{0}),\ \  b_{1}^{(2)}=\frac{a^{(2)}}{s_{1}^{(2)}}\left(\frac{q_{1}^{(1)}}{s_{1}^{(1)}}+\frac{\gamma_{(2)}^{d-2}}{a^{(1)2}}\right),\;\;\;b_{2}^{(2)}=0,\nonumber \\
	\mathcal{O}&(m^{-2}),\ \  c_{1}^{(2)}=-2(d-1)c-\frac{a^{(2)}q_{1}^{(1)}}{s_{1}^{(1)2}s_{1}^{(2)}}-\frac{2a^{(2)}\gamma_{(2)}^{d-2}}{a^{(1)2}s_{1}^{(1)}s_{1}^{(2)}},\;\;\;c_{3}^{(2)}=c+\frac{a^{(2)}\gamma_{(2)}^{d-2}}{a^{(1)2}(d-1)s_{1}^{(1)}s_{1}^{(2)}},\nonumber \\
	&c_{2}^{(2)}=2(d-1)c-\frac{2a^{(2)}\gamma_{(2)}^{d-2}}{(d-2)a^{(1)2}s_{1}^{(1)}s_{1}^{(2)}},\;\;\;c_{4}^{(2)}=-c+\frac{a^{(2)}\gamma_{(2)}^{d-2}}{a^{(1)2}(d-1)(d-2)s_{1}^{(1)}s_{1}^{(2)}},
\end{align}
where the  parameter $c$ is
\begin{align}\label{c}
	c=&-\frac{a^{(2)}q_{2}^{(1)}}{2(d-1)s_{1}^{(1)2}s_{1}^{(2)}}-\frac{a^{(2)}q_{1}^{(1)}s_{2}^{(1)}}{2(d-1)s_{1}^{(1)3}s_{1}^{(2)}}-\frac{a^{(2)}q_{1}^{(1)2}s_{2}^{(2)}}{2(d-1)s_{1}^{(1)2}s_{1}^{(2)3}}\nonumber \\
	&-\frac{a^{(2)}q_{1}^{(1)}s_{2}^{(2)}\gamma_{(2)}^{d-2}}{(d-1)a^{(1)2}s_{1}^{(1)}s_{1}^{(2)3}}-\frac{a^{(2)}s_{2}^{(2)}\gamma_{(2)}^{2(d-2)}}{2(d-1)a^{(1)4}s_{1}^{(2)3}}.
\end{align}
Substituting eqs.(\ref{S2Coefficients},\ref{c}) into eq.(\ref{S2Ansatz}), we get $S_{(2)}$
\begin{align}\label{S2Solution}
	S_{(2)\;j}^{\;i} =&  a^{(2)}\delta_{j}^{i}+\frac{a^{(2)}}{s_{1}^{(1)}m^{2}}(\frac{q_{1}^{(1)}}{s_{1}^{(1)}}+\frac{\gamma_{(2)}^{d-2}}{a^{(1)2}})\mathcal{P}_{1\;j}^{\;i}\nonumber\\
    +&\frac{1}{m^{4}}\left(c_{1}^{(2)}\mathcal{P}_{1\;k}^{i}\mathcal{P}_{1\;j}^{k}+c_{2}^{(2)}\mathcal{P}_{1}\mathcal{P}_{1\;j}^{i}+c_{3}^{(2)}\mathcal{P}_{1\;kl}\mathcal{P}_{1}^{\;kl}\delta_{j}^{i}+c_{4}^{(2)}\mathcal{P}_{1}^{\;2}\delta_{j}^{i}\right)+\mathcal{O}(m^{-6})\text{,}
\end{align}
where the parameters $c^{(b)}_n$ are given by eq.(\ref{S2Coefficients}).
Combining equations eqs.(\ref{h2Solution},\ref{S2Solution}),
we  express the metric $h_3$ in terms of the metric $h_1$
\begin{align}\label{h3Solution}
h_{3\;ij}
=&h_{2\;ik}(S_{(2)}^{\;2})_{\;j}^{k}\nonumber\\
=&a^{(1)2}a^{(2)2}\left[h_{1\;ij}+\frac{2(1+\frac{b_{1}^{(2)}s_{1}^{(1)}}{a^{(2)}})}{s_{1}^{(1)}m^{2}}\mathcal{P}_{1\;ij}+\text{\ensuremath{\frac{2\mathcal{P}_{1}\mathcal{P}_{1\;ij}}{s_{1}^{(1)3}m^{4}}}}(\frac{c_{2}^{(2)}s_{1}^{(1)3}}{a^{(2)}}-s_{2}^{(1)})\right]\nonumber\\
+&a^{(1)2}a^{(2)2}\cdot\frac{\mathcal{P}_{1\;ik}\mathcal{P}_{1\;j}^{k}}{a^{(2)2}s_{1}^{(1)3}m^{4}}\left(b_{1}^{(2)2}s_{1}^{(1)3}+2a^{(2)}s_{1}^{(1)2}(2b_{1}^{(2)}+c_{1}^{(2)}s_{1}^{(1)})+a^{(2)2}(s_{1}^{(1)}+2s_{2}^{(1)})\right)\nonumber\\
+&a^{(1)2}a^{(2)2}\cdot\frac{2s_{2}^{(1)}h_{1\;ij}}{s_{1}^{(1)3}m^{4}}\left(\frac{e_{2}(\mathcal{P}_{1})}{d-1}+\frac{s_{1}^{(1)3}}{a^{(2)}s_{2}^{(1)}}(c_{3}^{(2)}\mathcal{P}_{1\;kl}\mathcal{P}_{1}^{\;kl}+c_{4}^{(2)}\mathcal{P}_{1}^{\;2})\right).
\end{align}
Finally, combining eqs.(\ref{LintSToM},\ref{I6},\ref{L6},\ref{S2Coefficients},\ref{S2Solution},\ref{h3Solution}),
we obtain the higher derivative gravity with respect to two metrics $h_1$ and $h_6$
\begin{align}\label{I6R2}
	I_{6}^{HD}(h_{1},h_{6}) & =  m_{(1)}^{d-2}\int_{Q_{1}}d^{d}y \sqrt{|h_1|}\left[\Lambda^{(1)}+c_{\mathcal{R}}^{(1)}\mathcal{R}_{1}+\frac{c_{\mathcal{R}\mathcal{R}}^{(1)}}{m^{2}}\left(\mathcal{R}_{1}^{\;ij}\mathcal{R}_{1\;ij}-\frac{d}{4(d-1)}\mathcal{R}_{1}^{\;2}\right)\right]\nonumber \\
	& +  m_{(6)}^{d-2}\int_{Q_{2}}d^{d}y \sqrt{|h_6|}\left[\Lambda^{(6)}+c_{\mathcal{R}}^{(6)}\mathcal{R}_{6}+\frac{c_{\mathcal{R}\mathcal{R}}^{(6)}}{m^{2}}\left(\mathcal{R}_{6}^{\;ij}\mathcal{R}_{6\;ij}-\frac{d}{4(d-1)}\mathcal{R}_{6}^{\;2}\right)\right]\nonumber\\
	&+\mathcal{L}_{int\;16}(h_{1},h_{6})+\mathcal{O}(\mathcal{R}_{1}^{\;3})+\mathcal{O}(\mathcal{R}_{4}^{\;3}),
\end{align}
where the parameters are given by 
\begin{align}\label{I6R2Coefficient}
	\Lambda^{(1)}&=-2m^{2}(\alpha_{0}^{(1)}+a^{(1)d}\alpha_{0}^{(2)}),\;\;\;c_{\mathcal{R}\mathcal{R}}^{(1)}=\hat{c}_{\mathcal{RR}}^{(1)}+f_{1}+f_{2}+f_{3},\nonumber \\
	c_{\mathcal{R}}^{(1)}&=\hat{c}_{\mathcal{R}}^{(1)}+a^{(1)d}\left[\frac{\text{(}a^{(2)}\gamma_{(3)})^{d-2}}{a^{(1)2}}-\frac{(d-2)(a^{(2)}\alpha_{0}^{(2)}+b_{1}^{(2)}s_{1}^{(1)}\alpha_{1}^{(2)})}{(d-1)s_{1}^{(1)}a^{(2)}}\right],
\end{align}
and
\begin{align}
	f_{1}=&\frac{(a^{(1)}\gamma_{(2)})^{d-2}\left(\alpha_{2}^{(2)}\gamma_{(2)}^{d-2}-2s_{1}^{(2)}(a^{(2)}\gamma_{(3)})^{d-2}\right)}{a^{(1)2}s_{1}^{(2)2}},\nonumber\\
	f_{2}=&\frac{2\left(q_{1}^{(1)}\alpha_{2}^{(2)}+\frac{d-2}{d-1}s_{1}^{(2)}\alpha_{1}^{(2)}\right)(a^{(1)}\gamma_{(2)})^{d-2}-2s_{1}^{(2)}(q_{1}^{(1)}+s_{1}^{(2)})(a^{(1)}a^{(2)}\gamma_{(3)})^{d-2}}{s_{1}^{(1)}s_{1}^{(2)2}},\nonumber \\
	f_{3}=&\frac{a^{(1)d}}{a^{(2)}s_{1}^{(1)3}s_{1}^{(2)2}}\Bigg[a^{(2)}q_{1}^{(1)2}s_{1}^{(1)}\alpha_{2}^{(2)}+a^{(2)}s_{1}^{(2)2}\alpha_{0}^{(2)}(s_{1}^{(1)}-\frac{d-2}{d-1}s_{2}^{(1)}) \nonumber \\
	&+  2s_{1}^{(1)}s_{1}^{(2)}\alpha_{1}^{(2)}\left(a^{(2)}q_{1}^{(1)}+(d-2)cs_{1}^{(1)2}s_{1}^{(2)}\right) \Bigg].
\end{align}
Similarly, $\Lambda^{(6)},c_{\mathcal{R}}^{(6)}$ and $c_{\mathcal{RR}}^{(6)}$ can be obtained from  eq.(\ref{I6R2Coefficient}) by replacing indexes $(b)$ with $(N+1-b)$.
Remarkably, the multi-gravity with eight metrics ($N=8$)  eq.(\ref{I6R2}) yields the same curvature squared terms as the effective action eq.(\ref{IW}) on the brane. 
 It is expected that the results can be generalized to arbitrary $N$. This is a strong evidence for the equivalence between ghost-free multi-gravity and the higher derivative gravity eq.(\ref{IW}) on the brane.

\end{document}